\documentclass[9pt,twocolumn,twoside]{opticajnl}
\journal{opticajournal} 

\setboolean{shortarticle}{false}
\usepackage{color,soul}
\renewcommand{\hl}[1]{\textcolor{black}{#1}}

\usepackage{tabularx}
\usepackage{array}
\usepackage{booktabs}
\usepackage{lineno}

\title{From Solid-State Spin Defects to Fluorescent Proteins: A Perspective on Quantum Optical Sensing Platforms for Biomedical Applications}

\author[1]{Jyotiprakash Parhi}
\author[1]{Ayan Majumder}
\author[1,2,*]{Kasturi Saha}

\affil[1]{Department of Electrical Engineering, Indian Institute of Technology Bombay, Mumbai-400076, India}
\affil[2]{Center of Excellence for Quantum Information, Computing Science and Technology, Indian Institute of Technology Bombay, Mumbai-400076, India.}

\affil[*]{kasturis@ee.iitb.ac.in}

\begin{abstract}
Quantum optical sensing has evolved from isolated solid-state spin defect platforms to increasingly bio-integrated systems designed for biomedical applications. This review presents a perspective on this transition, spanning nitrogen-vacancy centers in diamond to emerging fluorescent protein–based spin qubits. Advances in diamond fabrication have enabled photonic cavities, nanomechanical resonators, and microstructured devices that enhance spin–photon interactions and room-temperature sensitivity. Exploration of spin defects in alternative materials further expands the sensing landscape. Parallel progress toward biological compatibility includes chip-based architectures for in vitro and in vivo studies, microfluidic integration, and minimally invasive nanodiamond probes capable of intracellular nanothermometry and nanorheometry. Improvements in surface functionalization have enhanced coherence and sensing reliability in complex environments. Collectively, these advances \hl{highlight the progress and remaining challenges toward application-oriented quantum sensing for the life sciences.}
\end{abstract}

\setboolean{displaycopyright}{false} 

\begin{document}

\maketitle

\section{Introduction}
\label{section-1}

Quantum sensing has emerged as one of the most powerful approaches for probing weak magnetic, electric, and thermal signals at the nanoscale. In contrast to classical sensors, quantum sensors are individual systems that exploit coherence, interference, and entanglement to transduce external perturbations into measurable observables~\cite{aslam2023quantum}. There is a generic protocol for performing measurements: initialization of the sensor, coupling to the signal, signal readout, and estimation~\cite{du2024single}. At its core, a quantum sensor operates by preparing a well-defined quantum state, allowing it to evolve under an interaction Hamiltonian $H=H_0 + H_{sig}$ ($H_0$ is the intrinsic Hamiltonian and $H_{sig}$ encodes coupling to external perturbations), and mapping the accumulated phase or population change onto a classical signal through measurement. Although different physical platforms exist, spin-based solid-state defects, most prominently the nitrogen-vacancy(NV) center in diamond and nanodiamonds, have become leading candidates for ambient and biological conditions~\cite{10.1063/5.0147469}.

The sensitivity of a spin-based quantum sensor can be expressed as: $\eta\propto\frac{1}{\sqrt{N}}\frac{1}{\gamma\sqrt{T}}$, with $N$ being the number of detected photons, $\gamma$ being the gyromagnetic ratio, and $T$ being the interrogation time~\cite{degen2017quantum}. Longer coherence time, efficient signal collection, coupled with higher signal contrast, contribute to enhanced sensitivity. In biological environments, fluctuating magnetic fields, surface noise, and charge instability often limit coherence. This makes the selection of protocol crucial, as the techniques relying on robust population measurement may outperform phase-sensitive methods. In contrast, $T_1$ relaxometry measures the longitudinal relaxation time, the timescale over which the spin population returns to thermal equilibrium. It is sensitive to the spectral density of the magnetic noise~\cite{degen2017quantum}, i.e., $\frac{1}{T_1}\propto S_B(\omega_0)$. Where $S_B(\omega_0)$ is the magnetic noise spectral density at frequency $\omega_0$. From a quantum sensing perspective, it is a way to probe stochastic, high-frequency magnetic fluctuations. In biological contexts, fluctuating magnetic species, such as free radicals or transition-metal ions, generate magnetic noise that shortens $ T_1$~\cite {mzyk2022relaxometry} due to their paramagnetic nature, providing insights into nanoscale spin dynamics.

\begin{figure}[htbp]
\centering
\includegraphics[width=\linewidth]{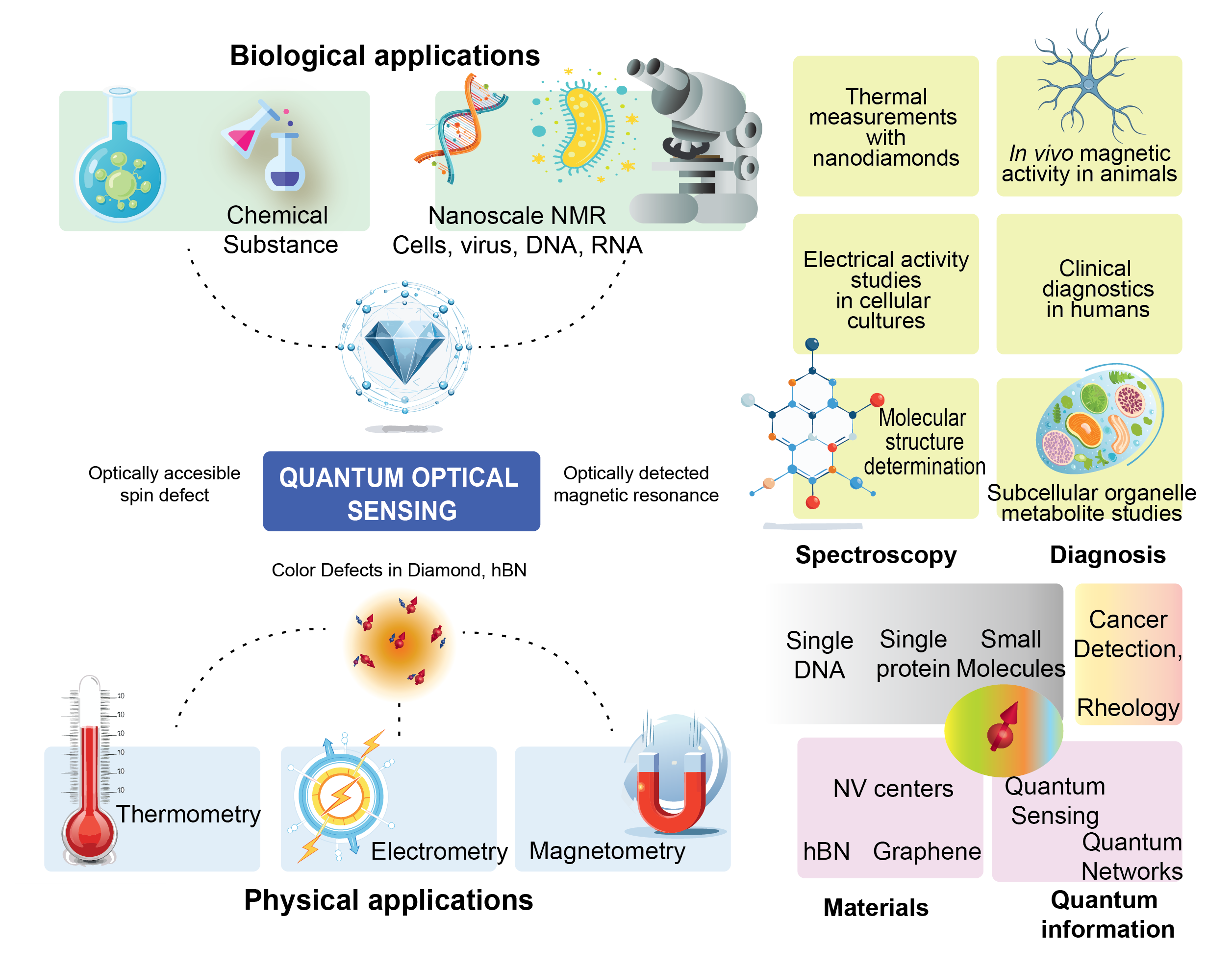}
\caption{Schematic overview of optically accessible spin defects as quantum sensor: from the nitrogen-vacancy center in diamond to hexagonal boron nitride. The applications range from physical sensing modalities to advanced spectroscopic techniques. The biocompatibility extends the capability from radical detection, chemical substance analysis, and rheology to nanoscale NMR of DNA, RNA, viruses, and in cellular environments. The schematic illustrates the versatility of quantum sensors in enabling a wide range of applications, spanning from biomedical research to quantum information technologies.}
\label{S1_Fig1}
\end{figure}

NV centers in diamond can be probed via optically detected magnetic resonance (ODMR), where spin-dependent fluorescence enables the detection of variations in magnetic field, electric field, strain, and temperature~\cite{sengottuvel2022wide,alam2024determining,dolde2011electric,szawiola2025thermal}. The zero-field splitting parameter $D$ of the spin-triplet ground state exhibits a temperature dependence arising from lattice expansion~\cite{10.1063/5.0147469} and electron--phonon interactions. Consequently, temperature changes produce measurable shifts in the spin-resonance frequency, with $\frac{dD}{dT}\approx -74~\mathrm{kHz/K}$~\cite{10.1063/5.0147469,degen2017quantum,du2024single}. Temperature sensing is therefore reduced to a precision frequency-metrology problem. A major challenge in biological environments is to distinguish temperature-induced shifts from those arising from strain, electric fields~\cite{dolde2011electric}, or magnetic fields, which requires careful calibration and drift compensation~\cite{du2024single,degen2017quantum}.

Double electron-electron resonance (DEER) provides a complementary approach for probing dipolar interactions between the NV sensor spin and nearby target spins. DEER extends the spin-echo sequence by introducing an additional microwave pulse that selectively flips the environmental spins~\cite{PhysRevB.104.224412}. For dipolar interactions, the coupling follows $H_{dip}\propto\frac{1}{r^3}$, where $r$ is the spin--spin separation. Flipping the nearby spins changes the local magnetic field experienced by the NV center, thereby modifying its phase accumulation and coherence. This provides spectral selectivity and information on nearby spin species, spin density, and coupling strength. However, DEER requires sufficiently long $T_2$ coherence times and precise microwave control~\cite{degen2017quantum,du2024single}, making it more challenging in noisy biological environments.

Translating quantum sensing protocols into biological environments requires several stringent conditions to be met. The sensor must function reliably in aqueous, room-temperature conditions typical of living systems. Surface-induced decoherence, charge instability, and environmental noise must be minimized to preserve long relaxation and coherence times. At the same time, the sensing platform must remain minimally invasive: microwave excitation and optical illumination should not introduce significant local heating or photodamage. \hl{Reliable calibration and efficient photon collection are also essential for quantitative measurements with a sufficient signal-to-noise ratio.} Within the broader quantum sensing framework, protocols such as $T_1$ relaxometry, ODMR, and DEER probe complementary observables; relaxation dynamics, transition frequencies, and phase accumulation, respectively. Despite probing different quantities, they share a common principle: controlled quantum evolution under an interaction Hamiltonian for sensitive biomedical measurements~\cite{degen2017quantum,du2024single,10.1063/5.0147469,10.1063/5.0170145,schirhagl2014nitrogen}. \hl{In complex biological settings, however, decoherence, surface noise, calibration stability, and biocompatibility remain important challenges.} \hl{This review provides a perspective on biological sensing with fluorescent quantum sensors, highlighting the underlying quantum sensing principles, representative experimental protocols, and the practical challenges that must be addressed for robust, quantitative, and biologically relevant measurements.}


\section{Optical and spin properties of solid-state color centers and fluorescent proteins}
\label{section-2}

Micron- and nanoscale sensors are essential for probing intracellular biochemical processes and resolving biomolecular structure at the single-molecule level~\cite{zhang2021toward,du2024single}. Solid-state quantum sensors have recently emerged as a powerful platform for such applications, enabling high-precision measurements and advanced quantum metrology~\cite{degen2017quantum,aslam2023quantum,das2024quantum}. Beyond expanding the limits of nanoscale detection, these systems hold strong potential for single-molecule studies and biomedical diagnostics, including monitoring disease-related cellular processes and drug interactions~\cite{chen2017diamond,Parhi2025.07.27.666994}. In this review, we discuss the spin and optical properties of leading solid-state quantum sensors—such as NV$^{-}$ and SiV centers in diamond, boron–vacancy centers in hBN, and emerging fluorescent-protein-based platforms—with emphasis on their operating principles and suitability for biological environments~\cite{10.1063/5.0170145,rizzato2023extending,feder2025fluorescent,abrahams2026quantum}.

\subsection{Optical and spin properties of solid-state color centers}

\begin{figure}[ht]
\centering
\includegraphics[width=\linewidth]{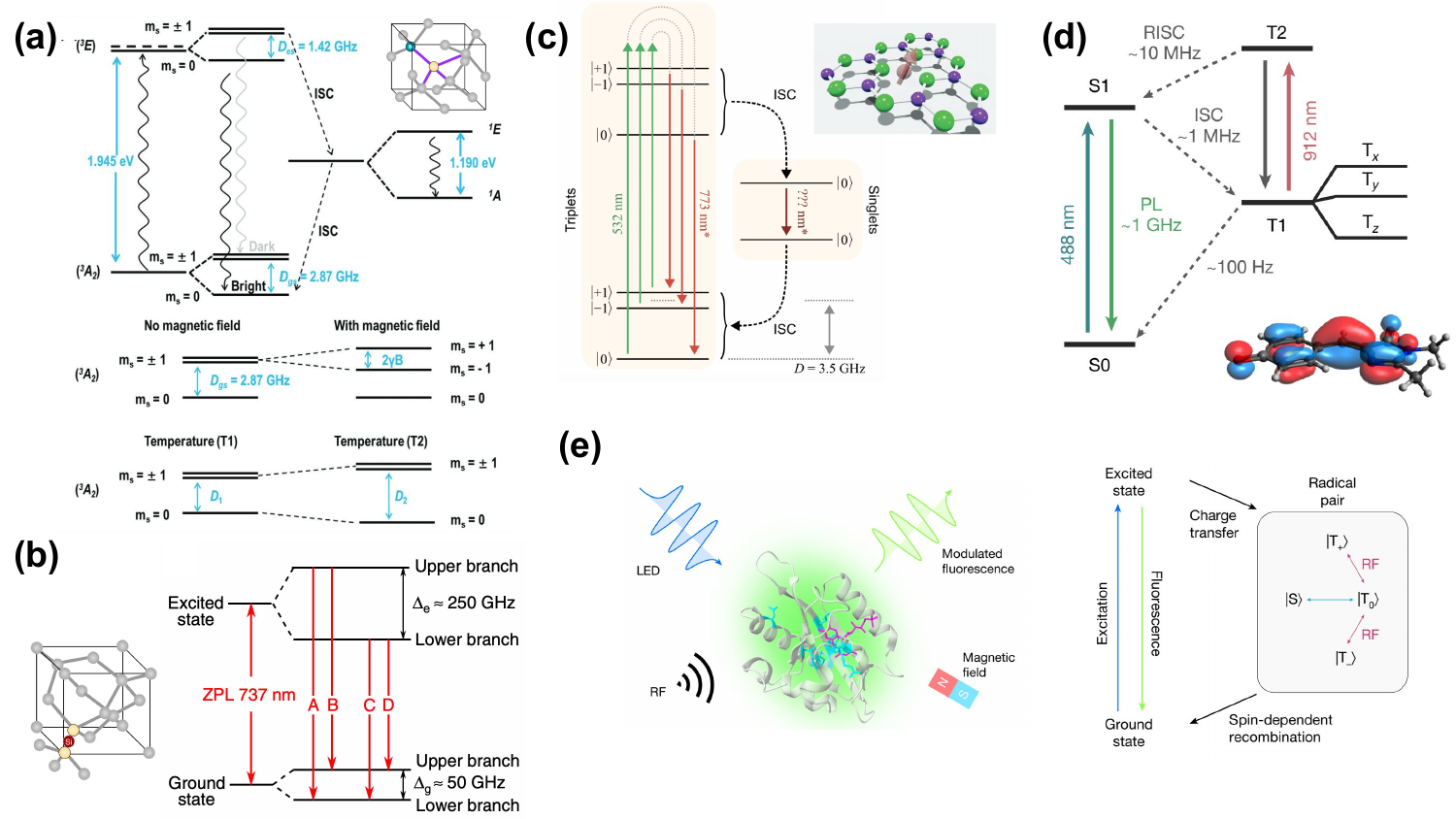}
\caption{(a) Electronic structure of the NV$^{-}$ center in diamond, showing triplet ground and excited states coupled by optical transitions and intersystem crossing (ISC) via singlets, enabling spin polarization and spin-dependent photoluminescence; magnetic-field and temperature effects are indicated (inset: atomic structure). Adapted from Refs.~\cite{suter2017single,10.1063/5.0170145}. (b) SiV$^{-}$ energy levels with spin–orbit-split ground and excited states, yielding four optical transitions (inset: split-vacancy structure along $\langle 111 \rangle$). Adapted from Ref.~\cite{Müller2014}. (c) Simplified level scheme of V$_\mathrm{B}^{-}$ in hBN with triplet ($S=1$) ground and excited states separated by $\sim$773~nm; 532~nm excitation and ISC enable spin polarization and readout (inset: boron-vacancy structure). Adapted from Refs.~\cite{robertson2025photophysics,mendelson2022coupling}. (d) EYFP energy diagram illustrating optically activated delayed-fluorescence (OADF) readout: 488~nm excitation and ISC populate $|T_1\rangle$, while a 912~nm pulse induces reverse ISC and spin-dependent delayed fluorescence (inset: $|T_1\rangle$ natural transition orbital). Adapted from Ref.~\cite{feder2025fluorescent}. (e) Radical-pair photocycle under a strong magnetic field, where triplet-born states ($|T_0\rangle$, $|T\pm\rangle$) undergo singlet–triplet mixing, producing magnetic-field-dependent reaction yields. Adapted from Ref.~\cite{abrahams2026quantum}.}
\label{S3_Fig1}
\end{figure}

Optically active defects in diamond, known as color centers, have become central to solid-state quantum optical sensing, with the negatively charged nitrogen–vacancy (NV$^{-}$) center emerging as the most versatile platform~\cite{mzyk2021diamond,suter2017single}. The NV defect, comprising a substitutional nitrogen atom adjacent to a vacancy, acts as a stable single-photon emitter with exceptional photostability and broadband fluorescence under 532~nm excitation~\cite{10.1063/5.0170145}. Its spin-triplet ($S=1$) ground state (Figure~\ref{S3_Fig1}(a)) enables optical initialization and spin-dependent readout via optically detected magnetic resonance (ODMR), allowing coherent microwave control at room temperature~\cite{suter2017single,levine2019principles}. The zero-field splitting and Zeeman interactions (Figure~\ref{S3_Fig1}(a)) render the NV center sensitive to temperature, strain, pressure, and magnetic fields, while spin-lattice relaxation ($T_1$) measurements provide access to magnetic noise and paramagnetic species in biological environments~\cite{barry2020sensitivity,mzyk2022relaxometry, modak2023few}. These combined optical and spin properties establish the NV center as a leading platform for nanoscale metrology and biosensing~\cite{degen2017quantum,zhang2021toward}. Furthermore, the biocompatibility of the diamond sensor extends to various applications and uses in biological environments~\cite{zhu2012biocompatibilit}.

Silicon vacancy (SiV$^{-}$) centers represent an attractive alternative platform for biosensing in diamond. The defect consists of a silicon atom positioned between two adjacent carbon vacancies along the $\langle 111 \rangle$ axis, forming a split-vacancy structure with $D_{3d}$ inversion symmetry (Figure~\ref{S3_Fig1}(b))~\cite{hepp2014electronic}. \hl{This inversion symmetry suppresses first-order sensitivity to electric-field noise, resulting in minimal spectral diffusion, narrow optical linewidths, and excellent spectral stability}~\cite{hepp2014electronic,Müller2014}. The SiV$^{-}$ center exhibits a characteristic four-line fine structure around 737~nm arising from split ground and excited states, with ~70-80\% of emission concentrated in the zero-phonon line (ZPL), reflecting weak electron-phonon coupling and a high Debye-Waller factor~\cite{hepp2014electronic,10.1063/5.0170145}. Notably, the ZPL position shifts linearly with temperature near room temperature, enabling sensitive optical thermometry based on lattice-expansion-induced modifications of the electronic structure~\cite{liu2022silicon,10.1063/1.5029904}. Beyond NV and SiV centers, additional diamond color defects are being explored for quantum sensing and bioimaging. Most emit in the visible and near-infrared-I (NIR-I, $\lambda < 1000$~nm) region~\cite{johnson2025nitrogen}, where tissue absorption and scattering can limit biological applications~\cite{turcheniuk2017biomedical,johnson2025nitrogen}. The nitrogen-vacancy-nitrogen (N$_2$V) center has recently gained attention, particularly its negatively charged state, N$_2$V$^{-}$, which is paramagnetic with spin $S=1/2$ and exhibits a well-resolved EPR spectrum~\cite{johnson2025nitrogen}. \hl{Notably,} $\mathrm{N}_2\mathrm{V}^{-}$ \hl{exhibits emission extending into the near-infrared-II (NIR-II, 1000--1700 nm) range}. This enables deeper tissue penetration compared to visible-emitting defects, highlighting its potential for deep-tissue bioimaging and biomedical sensing~\cite{johnson2025nitrogen,turcheniuk2017biomedical}.

Beyond diamond color centers, optically active spin defects in two-dimensional materials have emerged as promising nanoscale quantum sensors~\cite{robertson2025photophysics,rizzato2023extending}. The negatively charged boron–vacancy (V$_\mathrm{B}^{-}$) center in hBN (Figure~\ref{S3_Fig1}(c)) consists of a missing boron atom surrounded by three nitrogen atoms and hosts a spin-triplet ($S=1$) ground state with sublevels $|0\rangle$ and $|\pm 1\rangle$~\cite{robertson2025photophysics}. These states are separated by a zero-field splitting of $D \sim 3.47$~GHz and can be optically initialized and read out via spin-dependent photoluminescence under 532~nm excitation, with emission peaking near 850~nm. Similar spin-active defects have been realized in 10–50~nm hBN nanoparticles, exhibiting photostable emission (~820~nm), $T_1 \approx 17~\mu$s, and ODMR contrast of ~10\%~\cite{10.1063/5.0202699,robertson2025photophysics,rizzato2023extending}. Their atomic thickness, surface proximity, and optical robustness make V$*\mathrm{B}^{-}$ centers attractive for surface-sensitive and intracellular quantum sensing applications.

\hl{Beyond diamond and hBN, 4H-SiC has emerged as a complementary wide-bandgap platform for integrated quantum photonics. The 4H-SiCOI architecture combines CMOS-compatible thin-film processing with optically addressable spin defects and lithographically defined photonic structures. Coupling PL4 divacancies to tunable microring cavities enhances ZPL emission and ODMR contrast while preserving coherent spin control}~\cite{Lukin2020,Bao2025,Zhou2025SiC}. \hl{These developments make SiC promising for scalable spin-photon integration, although its biological sensing applications remain at an early stage.}

\subsection{Optical and spin properties of fluorescent proteins}

Current solid-state quantum sensors provide outstanding sensitivity, yet their use in biological environments is limited by device size, surface functionalization demands, and poor genetic compatibility~\cite{feder2025fluorescent,du2024single,degen2017quantum,abrahams2026quantum}. A molecular, genetically encodable qubit would overcome these constraints and enable in situ intracellular sensing. A promising candidate is enhanced yellow fluorescent protein (EYFP), whose fluorophore, embedded in a $\beta$-barrel, can be spin-polarized via intersystem crossing (ISC) under 488~nm excitation, populating a metastable triplet state $T_1$ (Figure~\ref{S3_Fig1}(d))~\cite{feder2025fluorescent}. Although direct readout from $T_1$ is limited by rapid spin depolarization and its long intrinsic lifetime, optical promotion to a higher triplet state ($T_2$) using a 912~nm pulse enhances reverse ISC, generating spin-dependent delayed fluorescence. This optically activated delayed-fluorescence (OADF) mechanism enables fast, efficient spin readout and significantly improves the sensitivity of EYFP-based quantum sensing~\cite{feder2025fluorescent}.

Beyond triplet-based fluorescent proteins, magneto-responsive fluorescent proteins (MFPs) rely on spin-correlated radical pairs (SCRPs) as the sensing element (Figure~\ref{S3_Fig1}(e))~\cite{henbest2004radio,abrahams2026quantum}. SCRPs form via photo-induced electron transfer, typically from aromatic residues such as tryptophan or tyrosine to flavin mononucleotide (FMN), and inherit a singlet ($|S\rangle$) or triplet ($|T\rangle$) spin configuration from their precursor state. Weak exchange coupling enables coherent singlet-triplet interconversion, governed primarily by hyperfine interactions and modulated by external magnetic fields. At low fields, all triplet sublevels mix efficiently with the singlet state, whereas increasing Zeeman splitting restricts mixing mainly to $|S\rangle$ and $|T_0\rangle$~\cite{abrahams2026quantum,finkler2021quantum,timmel2004study}. Because recombination pathways are spin-selective, singlets recombine directly, whereas triplets follow alternative chemical channels, making the reaction yield magnetic-field-dependent and forming the basis of radical-pair-based quantum sensing in biological systems.

\begin{table*}[t]
    \centering
    \caption{Properties and biological suitability of quantum sensing probes discussed in this review.}
    \label{tab:probe_comparison}

    \footnotesize
    \setlength{\tabcolsep}{3.5pt}
    \renewcommand{\arraystretch}{1.15}

    \begin{tabularx}{\textwidth}{@{}
        >{\raggedright\arraybackslash}p{0.14\textwidth}
        >{\raggedright\arraybackslash}p{0.29\textwidth}
        >{\raggedright\arraybackslash}p{0.20\textwidth}
        >{\raggedright\arraybackslash}X
        @{}}

        \toprule

        \textbf{Probe} &
        \textbf{Spin/readout properties and performance} &
        \textbf{Optical/MW requirements} &
        \textbf{Biological suitability} \\

        \midrule

        \textbf{Diamond NV$^{-}$\newline (bulk/shallow)}
       ~\cite{suter2017single,barry2020sensitivity,yu2024noninvasive}
        &
        $T_1 \sim$ 1--10 ms; $T_2$: $\mu$s--ms, depending on depth and protocol.
        Ensemble magnetometry: 9~pT~Hz$^{-1/2}$ in living-rat magnetocardiography.
        &
        Typically 532~nm excitation, red fluorescence, and GHz MW control.
        &
        High photostability and mature spin control; suited to biomagnetism,
        nanoscale magnetic resonance, and surface-proximal sensing. \\

        \addlinespace

        \textbf{Fluorescent nanodiamond (NV$^{-}$)}
       ~\cite{mzyk2022relaxometry,zvi2025engineering,gu2023simultaneous}
        &
        $T_1 \sim 0.1$--$0.4$~ms; spin-echo $T_2 \sim 1~\mu$s,
        with strong particle/surface dependence.
        Intracellular relaxometry, thermometry, and nanorheometry were demonstrated.
        &
        Visible excitation; MW for ODMR, whereas $T_1$ relaxometry can be MW-free.
        &
        Intracellular, mobile, and photostable; suited to local radical,
        temperature, magnetic-noise, and rheological sensing. \\

        \addlinespace

        \textbf{SiV$^{-}$ nanodiamond}
       ~\cite{10.1063/1.5029904,liu2022silicon}
        &
        ZPL near 738~nm.
        Temperature sensitivity $\sim521$~mK~Hz$^{-1/2}$.
        &
        All-optical readout; no MW required for ZPL thermometry.
        &
        Photostable near-IR probe for intracellular tracking and thermometry;
        strain/ZPL variability affects calibration. \\

        \addlinespace

        \textbf{hBN $V_{\mathrm{B}}^{-}$}
       ~\cite{rizzato2023extending,gao2023quantum}
        &
        $T_1 = 5.84\pm0.05~\mu$s; Hahn $T_2 = 58.5\pm0.4$~ns;
        DD extends $T_2$ to $\sim4.2~\mu$s.
        Gd$^{3+}$ detection: $\sim10^{-18}$~mol/image.
        &
        Typically 532~nm excitation, near-IR fluorescence, and GHz MW control.
        &
        Atomically thin geometry enables close sensor--sample proximity;
        suited to surface and liquid-phase sensing. \\

        \addlinespace

        \textbf{EYFP triplet spin qubit}
       ~\cite{feder2025fluorescent}
        &
        At 80~K: $T_1=141\pm5~\mu$s, Hahn $T_2=1.5\pm0.2~\mu$s,
        CPMG $T_2=16\pm2~\mu$s.
        RT ODMR contrast up to 8\%; sensitivity 2.7~mT~Hz$^{-1/2}$.
        &
        488-nm initialization, 912-nm OADF readout, and MW control.
        &
        Genetically encodable and targetable; limited by rapid RT spin relaxation
        and comparatively low sensitivity. \\

        \addlinespace

        \textbf{MagLOV / radical-pair protein}
       ~\cite{abrahams2026quantum}
        &
        Radical-pair sensor; conventional $T_1/T_2$ comparison not applicable.
        $\sim10\%$ single-cell ODMR contrast and
        $\sim26~\mu$T~Hz$^{-1/2}$ sensitivity at RT.
        &
        $\sim450$-nm excitation with RF spin driving.
        &
        Genetically encodable and functional at RT; suited to targeted sensing,
        but sensitive to the local photochemical environment. \\

        \bottomrule
    \end{tabularx}

    \vspace{2pt}

    \begin{minipage}{0.98\textwidth}
        \footnotesize
        \textit{Note:} RT: room temperature; DD: dynamical decoupling. The reported performance values predominantly correspond to ensemble or multi-emitter measurements and represent specific experimental conditions. Relaxation times and sensitivities depend on material quality, temperature, protocol, and local environment; different sensing modalities are therefore not directly comparable.
    \end{minipage}

\end{table*}


\hl{Table}~\ref{tab:probe_comparison}
\hl{compares representative measurements under their reported experimental conditions. Magnetic sensitivity, temperature sensitivity, and analyte detection limits describe different observables and should not be treated as equivalent performance metrics. Comparisons must also distinguish ensemble from single-cell measurements and low-temperature from room-temperature operation}
~\cite{barry2020sensitivity,rizzato2023extending,feder2025fluorescent,abrahams2026quantum}.

\hl{Probe selection should begin with the biological observable and the required access to the sample. Bulk NV ensembles are suitable for weak external magnetic-field measurements}
~\cite{barry2020sensitivity},
\hl{whereas fluorescent nanodiamonds provide intracellular access for local magnetic-noise sensing, with performance affected by particle and surface variability}
~\cite{mzyk2022relaxometry,zvi2025engineering}.
\hl{SiV nanodiamonds offer optical thermometry without microwave driving, while planar hBN places spin defects close to a sample interface}
~\cite{10.1063/1.5029904,gao2023quantum}.
\hl{Protein sensors provide genetic targeting, but their sensing mechanisms and operating conditions require separate consideration. In particular, the reported low-temperature EYFP coherence times do not establish equivalent room-temperature coherent control, and MagLOV radical-pair readout is not directly comparable to conventional NV coherence measurements}
~\cite{feder2025fluorescent,abrahams2026quantum}.
\hl{Platform selection must therefore consider sensitivity together with operating conditions, biological application, and integration requirements.}

\section{Emerging directions in quantum optical sensing for biomedical applications:}
\label{section-3}

\subsection{Probing single molecule dynamics}

\begin{figure}
\centering
\includegraphics[width=\linewidth]{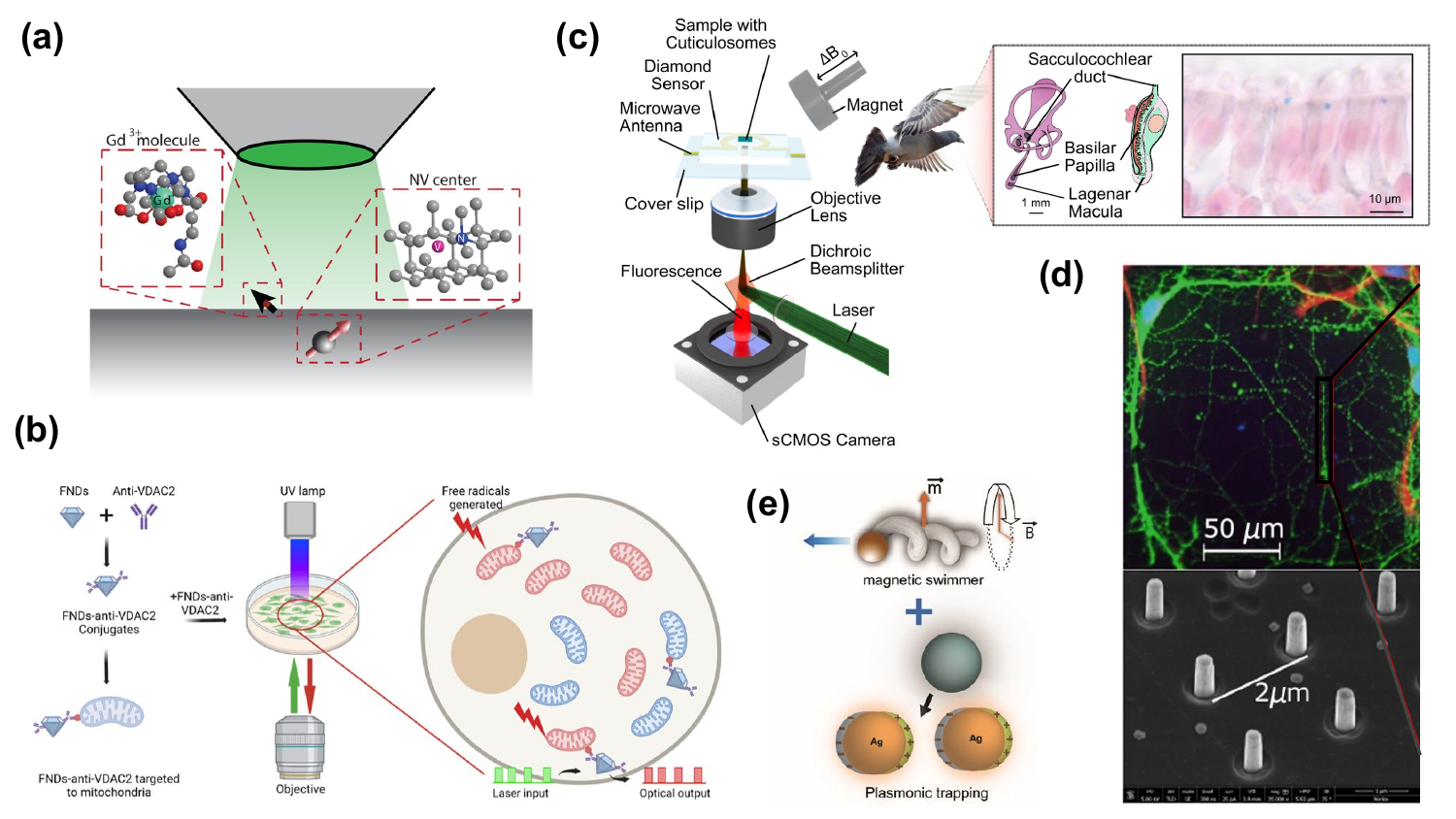}
\includegraphics[width=\linewidth]{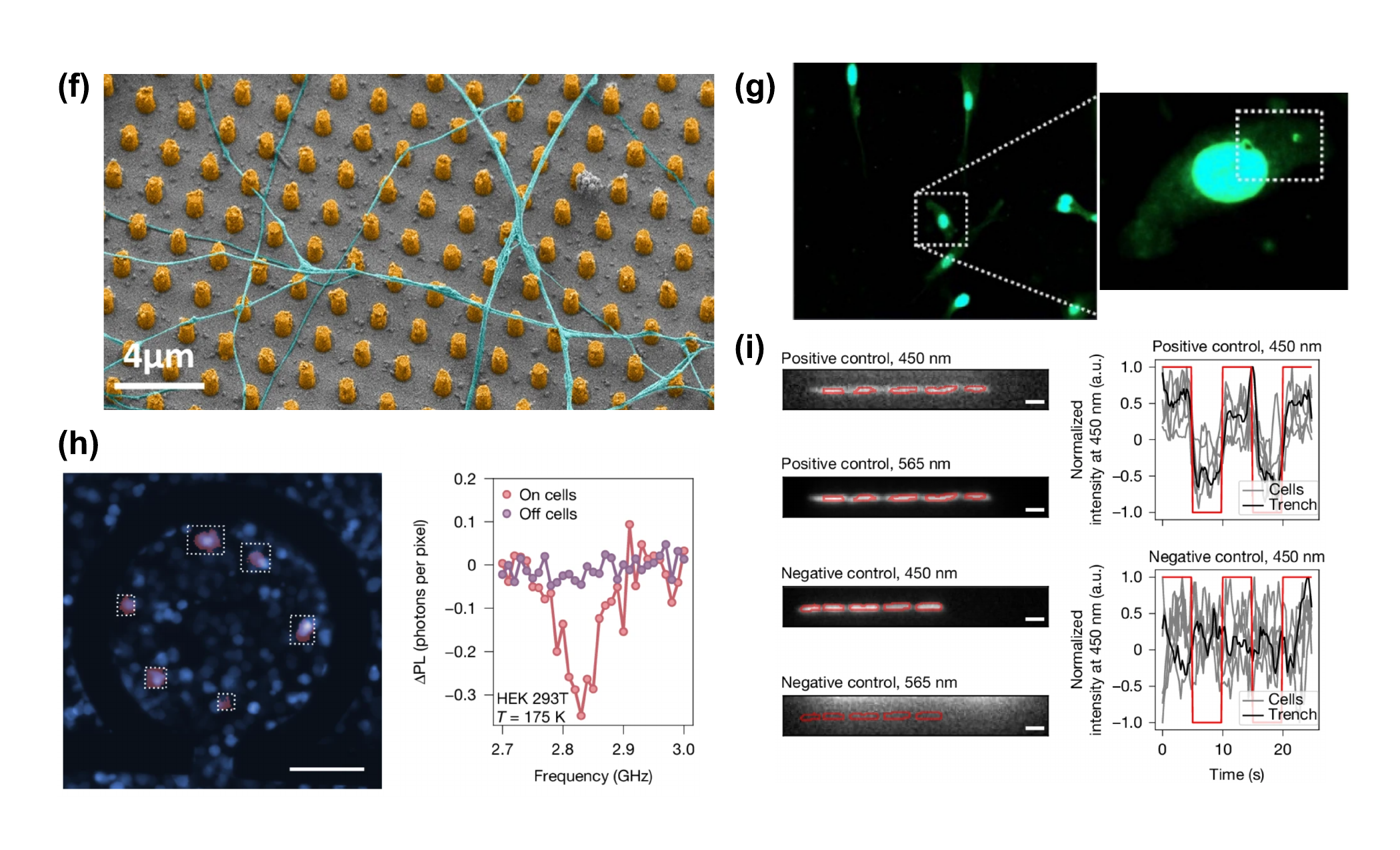}
\caption{(a) Single-molecule electron spin detection using a single shallow NV center in bulk diamond. Adapted from Ref.~\cite{doi:10.1021/nl502988n} (b)Schematic of NDs and the conjugation to anti-body VDAC2 for targeting mitochondria. Adapted from Ref.~\cite{doi:10.1021/acssensors.4c00118}. (c) (left) Overview of the design of the experiment. (right) Details of the biological sample of interest (pigeon ear and cochlea). Adapted from Ref.~\cite{doi:10.1073/pnas.2112749118} (d) (up) Confocal image of the stained neurons (green) grown on a bed of nanopillars. (down) SEM image of diamond nanopillars. Adapted from Ref.~\cite{hanlon2020diamond}.  (e) Magnetically actuated mobile nanotweezers used for controlled transport and manipulation of nanoscale objects, illustrating an auxiliary approach for positioning fluorescent nanodiamond probes. Adapted from Ref.~\cite{doi:10.1126/scirobotics.aaq0076}. (f) SEM images of primary hippocampal neurons plated on different diamond substrates. Adapted from Ref.~\cite{losero2023neuronal}. (g) Overlay of isolated fibroblast cells in the AF 488-T2 channel and cell nuclei stained with DAPI-T3. The white inset highlights the region containing single-crystal diamond nanoneedles (SCDNNs) and part of the cell. Adapted from Ref.~\cite{quarshie2025diamond}. (h) (left) A wide-field fluorescence image of HEK293T cells expressing cytosolic EYFP (blue) and OADF (red). Scale bar, $100\mu m$. (right) ODMR signal averaged over the bright pixels (red, on cells) and dark pixels (purple, off cells). Adapted from Ref.~\cite{feder2025fluorescent} (i) (left) Single trenches, with cells circled in red. Scale bar, $2\mu m$. (right) Time series of the 450-nm illumination fluorescence for the individual cells depicted in (left) over time, magnetic field 10mT (red), each cell(grey), average over all cells(black). Adapted from Ref.~\cite{abrahams2026quantum}}
\label{S4_Fig1,2}
\end{figure}


The investigation of the dynamics and the interactions of biomolecules at the single-cell level has emerged as a new paradigm of research, enabling a deeper understanding of biological and chemical processes within the cellular environment. Spin-based measurement techniques can overcome the challenges encountered by optical and electrochemical approaches~\cite{doi:10.1073/pnas.2509102122} in the direct interrogation of single biomolecules under ambient conditions. In a seminal demonstration, Li \textit{et al.}~\cite{doi:10.1073/pnas.2509102122} employed NV $T_1$ relaxometry to monitor biomolecular binding events through magnetic noise spectroscopy. The protein 'A' is immobilized, and when the spin-labeled protein 'B' comes near 'A', there will be an acceleration of the NV spin relaxation rate as a consequence of magnetic spin fluctuation from proximal paramagnetic levels. Through meticulous engineering of shallow NV centers and precise surface functionalization, they have achieved micrometer-scale detection in bulk diamond with a significant sensitivity enhancement of $4.5\times$ due to amplitude-weighted analysis of biexponential relaxation. Additionally, they have accomplished nanometer-scale detection within individual spin levels in nano-pillars, revealing distinct relaxometric signatures that correspond to the molecular binding and unbinding kinetics. Sushkov \textit{et al.}~\cite{doi:10.1021/nl502988n} employed shallow NV centers as a non-invasive and highly sensitive probe for the magnetic sensing of the individual paramagnetic molecules. They have attached the target molecules to the diamond surface covalently, and the readout for magnetic sensing is done optically, making it an all-optical nanoscale magnetometer (Figure~\ref{S4_Fig1,2}(a)). Due to their proximity to the surface, shallow NV centers are capable of detecting magnetic fluctuations generated by the electronic spin of a single paramagnetic molecule while maintaining favorable spin coherence and optical properties~\cite{doi:10.1021/nl502988n}. They integrated a three-step co-localization technique, combining optical fluorescence, gold nanoparticles, and atomic force microscopy, to isolate a single molecule within a sensing radius of 12~nm~\cite{doi:10.1021/nl502988n}, thereby establishing a label-free platform capable of measuring single-molecule dynamics.

Advancements in structural biology are crucial for understanding the conformational changes of individual molecules. This knowledge provides clear insights into the dynamics and interactions of organic molecules with their environment and proteins. \hl{C. Munuera-Javaloy \textit{et al.}}~\cite{Munuera-Javaloy2022} \hl{introduced a quantum-enhanced protocol for detecting conformational transitions in single molecules under ambient conditions. They considered shallow NV centers to detect magnetic interactions between two nitroxide electron-spin labels attached to a target molecule, such as a protein. An engineered double electron--electron resonance sequence was used to resolve small energy shifts arising from the coupling between the spin labels. By combining the simulated NV response with Bayesian inference, they showed that different molecular conformations and inter-label distances could be distinguished under realistic ambient conditions. This approach therefore provides a possible route toward probing single-molecule conformational dynamics, although experimental validation in living cells remains to be demonstrated.}

At the single-cell level, the mechanical interaction between the cell and the substrate is pivotal for understanding tissue repair, tumor metastasis, and morphogenesis~\cite{doi:10.1126/sciadv.adi5300}. In the advancement of mechanobiology, Xu \textit{et al.}~\cite{doi:10.1126/sciadv.adi5300} developed quantum-enhanced diamond molecular tension microscopy (QDMTM), a platform that leverages the quantum nature of electronic spins of NV centers in diamond to quantify cellular forces with unprecedented precision. By embedding shallow NV centers around \hl{5 $\pm$ 2~nm} from the diamond surface, the authors designed a sensitive magnetic interface, which is capable of detecting piconewton-scale mechanical pull from the living cell. The strength of this approach lies in its molecular design: cells are tethered to the diamond surface via a force-sensitive polymer chain that acts as a spring, with one end attached to the diamond and the other carrying a magnetic $Gd^{+3}$ tag. When the cell exerts force on a biological ligand attached to the spring, the resulting displacement of the magnetic tag generates magnetic fluctuations around the NV center, which in turn alter the spin relaxation time. \hl{Xu et al. established a framework for investigating mechanical signals by mapping integrin-based adhesion forces.} 


\subsection{Intracellular dynamics}

\hl{Breast cancer is the most frequently diagnosed cancer among women worldwide, accounting for approximately 2.3 million new cases in 2022, representing 11.6\% of all newly diagnosed cancers worldwide}~\cite{Bray2024} \hl{or 23.8\% of newly diagnosed cancers in women}~\cite{zhang2025global}. Wu \textit{et al.}~\cite{https://doi.org/10.1002/adfm.202514294} developed the first theranostic platform that combines quantum sensing and pH-sensitive drug delivery. This platform is designed to treat triple-negative breast cancer(TNBC), which uses hyper-branched polyglycerol (HPG) and 3-(4-methyl-2,5-dioxo-2,5-dihydrofuran-3-yl)propanoic acid (DMA) to conjugate the therapeutic drug diazoxide (DZX) for TNBC to the surface of NDs. When the functionalized NDs are internalized via endocytosis to endo/-lysosomal compartments, the mildly acidic environment(pH~6.5) causes cleavage of the acid linkage, precisely releasing the DZX drug in situ. This platform enables simultaneous imaging as well as sensing with $T_1$ relaxometry in the vicinity of the drug release. The release of DZX triggers the opening of ATP-sensitive potassium ($mitoK_{ATP}$) channels in mitochondria, which elevates the mitochondrial reactive oxygen species(ROS) production~\cite{https://doi.org/10.1002/adfm.202514294}. This simultaneously lowers the free radical generation in the endolysosomal regions. Wu \textit{et al.} demonstrated in vitro release of the DZX drug over a time period in different pH media. This platform can monitor intracellular redox homeostasis, which reveals organelle-resolved insights into drug mechanisms and sensing with relaxometry. \hl{Fan \textit{et al.}}~\cite{doi:10.1021/acssensors.4c00118} \hl{used $T_1$ relaxometry to monitor UVB-induced free-radical generation near mitochondria in keratinocytes using anti-VDAC2-functionalized nanodiamonds} (Figure~\ref{S4_Fig1,2}(b)). \hl{The targeted probes enabled changes in mitochondrial magnetic noise following UVB exposure to be detected close to the site of radical generation.}

\hl{Lin \textit{et al.}}~\cite{10.1093/hropen/hoaf007} \hl{used quantum sensing to investigate the metabolic response of human granulosa cells to follicle-stimulating hormone}. \hl{They showed that the hormone induces rapid free-radical generation without substantially compromising antioxidant defense or causing oxidative stress, suggesting a regulated signaling role for radicals in the reproductive process. Zhang \textit{et al.}}~\cite{doi:10.1021/acs.nanolett.4c02409} \hl{applied $T_1$ relaxometry to primary bronchial epithelial cells to study responses associated with chronic obstructive pulmonary disease (COPD). COPD-derived cells showed higher radical levels and a stronger response to cigarette-smoke extract than healthy cells. In a separate study, $T_1$ relaxometry was performed on fixed glioblastoma cells treated with increasing doses of temozolomide (TMZ), where a dose-dependent decrease in relaxation time indicated persistent radical-associated changes following treatment}~\cite{Parhi2025.07.27.666994}. \hl{Together, these studies demonstrate the potential of $T_1$ relaxometry to monitor disease- and treatment-associated changes in intracellular radical levels}.

\hl{Localized temperature changes within cells can influence cellular function, yet remain difficult to measure directly. Wu \textit{et al.}}~\cite{https://doi.org/10.1002/adma.202517076} \hl{developed a multifunctional nanodiamond platform that combines localized heating with real-time NV-based thermometry to study macrophage polarization. FND-CR particles accumulated in macrophage lysosomes, where NIR irradiation induced localized hyperthermia. The same particles measured temperature through ODMR and oxidative stress through $T_1$ relaxometry. At an irradiation power of $0.6~\mathrm{W/cm^2}$, the lysosomal temperature increased by approximately 24-25~$^\circ$C, reaching about 46-50~$^\circ$C. Reducing the irradiation power to $0.3~\mathrm{W/cm^2}$ produced a temperature rise of approximately 11~$^\circ$C after 15 minutes and did not induce macrophage polarization. The localized hyperthermia was accompanied by oxidative stress and transcriptional changes that drove the cells toward a pro-inflammatory (M1-like) phenotype}~\cite{https://doi.org/10.1002/adma.202517076}.

\hl{In contrast, Consiglio \textit{et al.}}~\cite{10.1021/acsanm.5c05655} \hl{used FNDs to resolve smaller metabolic temperature changes in glioblastoma cells, reporting an increase of approximately 1.7~$^\circ$C following mitochondrial uncoupling with FCCP. Their study addressed the challenge of detecting small intracellular temperature variations through careful ODMR calibration and statistical analysis. More recently, Sow \textit{et al.}}~\cite{sow2025millikelvin} \hl{developed a pulsed nanodiamond thermometry protocol that achieved an intracellular temperature precision of approximately 100~mK while reducing artifacts associated with surface electric-field fluctuations}. \hl{Under their experimental conditions, they observed no significant temperature change following metabolic stimulation. Together, these studies illustrate the range of NV-based thermometry, from resolving subtle metabolic temperature changes to generating and monitoring localized subcellular hyperthermia.}

\hl{Beyond temperature sensing, nanodiamonds have also been used to probe intracellular mechanical properties. Gu \textit{et al.}}~\cite{gu2023simultaneous} \hl{combined NV-based thermometry with nanodiamond tracking to perform simultaneous nanothermometry and nanorheometry in living cells. More recently, Singam \textit{et al.}}~\cite{Singam2026} \hl{used holotomographic tracking of NV-containing nanodiamonds to quantify effective intracellular viscosity in living cells. In the latter case, the viscosity readout is obtained from particle motion rather than NV spin dynamics, but it further demonstrates the use of nanodiamonds as intracellular biophysical probes.}


\subsection{Tissue and complex biological systems}

\hl{In recent years, quantum sensing has progressed from single molecules and cells to tissues and more complex multicellular systems. Zhang \textit{et al.}}~\cite{doi:10.1073/pnas.2317921121} \hl{used precision-cut mouse liver slices to demonstrate $T_1$ relaxometry for free-radical detection in living tissue. The tissue architecture was preserved while oxidative stress was induced using ethanol, resulting in increased radical levels and a corresponding decrease in $T_1$. In contrast, treatment with the antioxidant L-ascorbic acid increased $T_1$, indicating a reduction in the radical load. This study demonstrated that nanodiamond relaxometry can monitor changes in oxidative stress in a multicellular tissue environment despite challenges such as tissue autofluorescence and reduced optical penetration.}

\hl{The versatility of quantum sensing also extends to questions in sensory biology, including avian magnetoreception. de Gille \textit{et al.}}~\cite{doi:10.1073/pnas.2112749118} \hl{performed wide-field quantum magnetic imaging of iron-rich organelles known as cuticulosomes in thin sections of the pigeon cochlea (Figure~\ref{S4_Fig1,2}(c)). Using ODMR microscopy, they measured the stray magnetic fields generated by individual cuticulosomes under an applied magnetic field and compared the measurements with an analytical model to estimate their magnetic susceptibility. The measured susceptibility ($0.12\pm0.05$) was insufficient to support a particle-based magnetoreceptor mechanism and was found to lie between that of ferrihydrite and magnetite. Quantum relaxometry was further used to probe superparamagnetic components, with two of the ten cuticulosomes showing clear changes in the $T_1$ relaxation rate. These results indicate the presence of both superparamagnetic and ferrimagnetic components within the organelles and support alternative physiological roles such as iron storage or stereocilia stabilization. The study demonstrates how quantum magnetic imaging can characterize endogenous magnetic structures directly within biological tissue.}

\hl{In another study, Mzyk \textit{et al.}}~\cite{https://doi.org/10.1002/adfm.202528248} \hl{used nanodiamond-based quantum sensing to monitor reactive oxygen/nitrogen species and endolysosomal iron dynamics in human cardiac fibroblasts. They investigated how substrate stiffness and transforming growth factor-beta (TGF-$\beta$) influence these processes at different stages of endolysosomal maturation. The $T_1$ response varied with mechanical and biochemical stimulation, reflecting condition-dependent changes in free-radical and paramagnetic-iron levels. These changes were also influenced by cell passage number and the transition toward a myofibroblast-like phenotype. The study demonstrates that FND relaxometry can resolve redox and iron-associated changes in a cellular model of cardiac fibrosis and provides a useful platform for studying how mechanical and biochemical cues influence endolysosomal activity. Further validation in tissue and animal models will be required before such signatures can be considered predictive biomarkers of fibrotic progression.}

Two more studies underscore the applicability of the diamond quantum sensor in physiologically relevant and electrically active signals in biological systems. Nie \textit{et al.}~\cite{doi:10.1021/acs.nanolett.1c03021} extend the $T_1$ relaxometry to primary human dendritic cells, enabling label-free monitoring of intracellular radical dynamics during immune activation, giving insight into translational relevance beyond cell lines. Complementing the biochemical sensing, Barry \textit{et al.}~\cite{doi:10.1073/pnas.1601513113} reported the magnetic detection of single-neuron action potentials \hl{using an ensemble of NV centers in a bulk diamond sensor}, achieving millisecond temporal resolution sufficient to resolve neural firing events. They demonstrated that NV-based platforms can access electrophysiological signals in intact neurons by optically detecting tiny magnetic fields.

\hl{Yu \textit{et al.}}~\cite{yu2024noninvasive} \hl{demonstrated high-sensitivity, room-temperature magnetocardiography (MCG) using NV centers in diamond and non-invasively detected the cardiac magnetic signal of a living rat. Magnetic flux concentration was used to enhance the weak biomagnetic signal, achieving a sensitivity of $9~\mathrm{pT~Hz^{-1/2}}$. This study extends NV magnetometry to physiologically relevant in vivo measurements. Complementing this work, Bathla \textit{et al.}}~\cite{Bathla:26} \hl{demonstrated NV-based wide-field magnetic microscopy for dynamic 3D imaging and localization of time-varying current sources. Using currents designed to emulate neuronal activity, they reconstructed magnetic-field maps with micron-scale spatial and millisecond-scale temporal resolution. This approach enables the localization of active current sources across space and time and extends NV magnetometry toward more complex biological signals.}

\hl{Together, these studies illustrate the increasing biological complexity in which diamond quantum sensors have been tested, from cultured cells to tissue and living animals. To distinguish proof-of-principle studies from more advanced biological demonstrations,} Table~\ref{tab:biomedical_maturity} \hl{summarizes the highest stage demonstrated for each probe and the main barrier to further translation.}


\begin{table*}[t]
    \centering
    \caption{Biomedical maturity and principal translational barriers of quantum sensing probes.}
    \label{tab:biomedical_maturity}

    \footnotesize
    \setlength{\tabcolsep}{3.5pt}
    \renewcommand{\arraystretch}{1.15}

    \begin{tabularx}{\textwidth}{@{}
        >{\raggedright\arraybackslash}p{0.15\textwidth}
        >{\raggedright\arraybackslash}p{0.16\textwidth}
        >{\raggedright\arraybackslash}p{0.29\textwidth}
        >{\raggedright\arraybackslash}X
        @{}}

        \toprule

        \textbf{Probe} &
        \textbf{Most advanced biomedical stage} &
        \textbf{Representative evidence} &
        \textbf{Main translational barrier} \\

        \midrule

        \textbf{Diamond NV$^{-}$\newline (bulk/shallow)}
       ~\cite{yu2024noninvasive,aslam2023quantum}
        &
        \textbf{Animal / in vivo}\newline
        (external sensing)
        &
        Non-invasive cardiac magnetometry in a living rat.
        &
        Sensor–tissue distance, weak biomagnetic signals, and system complexity. \\

        \addlinespace

        \textbf{Fluorescent nanodiamond\newline (NV$^{-}$)}
       ~\cite{doi:10.1073/pnas.2317921121,Hamoya2024,shanahan2025q}
        &
        \textbf{Animal / in vivo}\newline
        (localized sensing)
        &
        Ex vivo liver radical sensing; intravital rat thermometry.
        &
        Optical penetration, motion, MW heating, biodistribution, and long-term safety. \\

        \addlinespace

        \textbf{SiV$^{-}$ nanodiamond}
       ~\cite{liu2022silicon}
        &
        \textbf{Cellular / in vitro}
        &
        Intracellular imaging, tracking, and thermometry.
        &
        ZPL calibration, targeted delivery, and lack of tissue/in-vivo validation. \\

        \addlinespace

        \textbf{hBN $V_{\mathrm{B}}^{-}$}
       ~\cite{Shimomura2026,Vlassiouk2026}
        &
        \textbf{Cellular proof-of-concept}
        &
        Intracellular ODMR after uptake by HeLa cells.
        &
        Defect reproducibility and progression to tissue/in-vivo sensing. \\

        \addlinespace

        \textbf{EYFP triplet spin qubit}
       ~\cite{feder2025fluorescent}
        &
        \textbf{Cellular proof-of-principle}
        &
        Mammalian-cell control at 175 K; ODMR in \textit{E.~coli} at room temperature.
        &
        Rapid room-temperature spin relaxation, photobleaching, and mammalian/in-vivo validation. \\

        \addlinespace

        \textbf{MagLOV / radical-pair protein}
       ~\cite{abrahams2026quantum}
        &
        \textbf{Cellular proof-of-principle}
        &
        Single-cell ODMR in \textit{E.~coli} at room temperature.
        &
        Validation in higher biological systems and environmental calibration. \\

        \bottomrule
    \end{tabularx}

    \vspace{2pt}

    \begin{minipage}{0.98\textwidth}
        \footnotesize
        \textit{Note:} Biomedical maturity is classified conservatively as
        proof-of-principle $\rightarrow$ cellular/in vitro $\rightarrow$
        ex vivo tissue $\rightarrow$ animal/in vivo $\rightarrow$ clinical/human.
        None of the platforms listed here has yet reached validated clinical/human
        quantum sensing in the cited studies.
    \end{minipage}

\end{table*}

\hl{Moving from cells to living tissue introduces limitations that are not considered by sensor sensitivity alone. Optical scattering and absorption reduce excitation and fluorescence collection, while physiological motion can introduce measurement instability. Microwave-driven protocols also require control of local heating, and nanoparticle probes raise additional questions of biodistribution and long-term biocompatibility }~\cite{Hamoya2024,shanahan2025q}.

\hl{A further challenge is separating different environmental contributions to the same sensor response. Temperature and magnetic field can both shift ODMR resonances, whereas fluctuating paramagnetic species primarily affect $T_1$; chemically transduced quantities such as pH can introduce additional cross-sensitivity. Simultaneous temperature and magnetic-field measurements have been demonstrated using nanodiamond ODMR}~\cite{Sarkar2026Dual}, \hl{while Bayesian and machine-learning approaches can jointly infer magnetic field and dephasing from noisy NV measurements}~\cite{Santagati2019}. \hl{Reliable separation of temperature, pH, and magnetic noise in complex in-vivo environments, however, remains unresolved.}

\subsection{Nanophotonic and integrated platforms for quantum biosensing}

\hl{Diamond nanostructures provide a route to improve optical readout while controlling the interface between the quantum sensor and the biological sample. Nanophotonic geometries can enhance photon collection from embedded color centers}~\cite{chen2017diamond,zhu2023multicone,rani2020recent}. \hl{In particular, NV-containing nanopillar arrays have been explored for neuronal sensing, where the nanostructures bring shallow NV centers close to electrically active cells while also providing a substrate for neuronal growth. Neuron-compatible arrays support viable hippocampal networks} (Figure~\ref{S4_Fig1,2}(f)), \hl{while modelling of the neuron--nanopillar interface illustrates their potential for local electric-field sensing} (Figure~\ref{S4_Fig1,2}(d))~\cite{hanlon2020diamond,losero2023neuronal}. \hl{Similarly, single-crystal diamond nanoneedles containing SiV centers have been interfaced with fibroblasts with low cytotoxicity and clear optical detectability} (Figure~\ref{S4_Fig1,2}(g))~\cite{quarshie2025diamond}. \hl{These geometries provide close sensor--sample interfaces, although reproducible defect placement and device uniformity remain important for quantitative sensing.}

\hl{Genetic expression offers another route to biological integration. Feder \textit{et al.}~\cite{feder2025fluorescent} demonstrated EYFP spin-dependent optical readout in mammalian cells at 175 K} (Figure~\ref{S4_Fig1,2}(h)), \hl{while Abrahams \textit{et al.}~\cite{abrahams2026quantum} demonstrated room-temperature spin-resonance readout using an engineered radical-pair protein} (Figure~\ref{S4_Fig1,2}(i)). \hl{Solid-state probes provide complementary approaches. SiV-containing nanodiamonds have been used for live-cell imaging and all-optical thermometry through temperature-dependent shifts of the $\sim$738-nm ZPL}~\cite{liu2022silicon}. \hl{Near-surface V$_{\mathrm{B}}^{-}$ defects in hBN provide another geometry for sensing paramagnetic species close to liquid interfaces}~\cite{gao2023quantum}.

\hl{Controlled positioning of nanoscale probes is another practical requirement for localized measurements. Opto-thermoelectric trapping has enabled manipulation of individual fluorescent nanodiamonds using plasmonic structures}~\cite{shukla2023opto}, \hl{while mobile nanotweezers combining magnetic actuation with plasmonic trapping can transport and position nanoscale objects, including fluorescent nanodiamonds} (Figure~\ref{S4_Fig1,2}(e))~\cite{doi:10.1126/scirobotics.aaq0076}. \hl{These approaches are not quantum sensing modalities themselves, but can provide supporting tools for positioning and manipulating quantum probes in fluidic or biological environments.}

\hl{At the device level, quantum sensing is increasingly being combined with electronic, photonic, and microfluidic components. CMOS-integrated NV sensors and photonic sensor arrays provide routes toward compact and multi-site readout}~\cite{Kim2019,Ibrahim2021,Pogorzelski2024,Weng2026}, \hl{while microfluidic platforms enable quantum sensing in controlled liquid environments and flowing droplets}~\cite{sarkar2024high,10.1039/d2lc00874b,fujiwara2023diamond}. \hl{Q-BiC further integrates microwave delivery, thermal control, and biological interfacing within a compact sensing platform}~\cite{shanahan2025q}. \hl{Related approaches are emerging in hBN and 4H-SiC, where planar spin defects and lithographically defined photonic structures provide additional routes toward scalable devices}~\cite{gao2023quantum,Vlassiouk2026,Lukin2020,Bao2025,Bao2026Deterministic}.

\hl{Moving from individual demonstrations to scalable quantum-biosensing platforms requires reproducible fabrication and integration. For diamond architectures, controlled defect placement and sensor uniformity remain important, while chip-based systems additionally require reliable optical, microwave, electronic, and fluidic integration. Emerging hBN and 4H-SiC platforms offer complementary routes toward scalable devices, although defect engineering and biological interfacing remain important challenges. These engineering considerations are summarized in Table}~\ref{tab:engineering_bottlenecks}.

Collectively, these developments show the increasing integration of quantum sensing platforms with biological systems. Their performance, however, remains closely linked to material- and interface-level limitations, as discussed in the following section.


\begin{table*}[t]
    \centering
    \caption{Quantum sensing platforms and key engineering bottlenecks.}
    \label{tab:engineering_bottlenecks}

    \footnotesize
    \setlength{\tabcolsep}{4pt}
    \renewcommand{\arraystretch}{1.12}

    \begin{tabularx}{\textwidth}{@{}
        >{\raggedright\arraybackslash}p{0.22\textwidth}
        >{\raggedright\arraybackslash}X
        >{\raggedright\arraybackslash}X
        @{}}

    \toprule
    \textbf{Platform} &
    \textbf{Demonstration} &
    \textbf{Engineering bottleneck} \\
    \midrule

    Diamond nanopillar arrays (NV$^{-}$)
   ~\cite{hanlon2020diamond,losero2023neuronal,rani2020recent}
    &
    Lithographically defined nanopillar arrays with shallow NV centers; neuronal compatibility demonstrated.
    &
    Controlled NV occupancy and positioning across large arrays. \\

    Diamond nanoneedles (SiV$^{-}$)
   ~\cite{quarshie2025diamond}
    &
    SiV-containing nanoneedles interfaced with fibroblasts.
    &
    Control of needle dimensions and defect distribution; cellular spin sensing remains to be demonstrated. \\

    CMOS-/photonic-integrated NV devices
   ~\cite{Kim2019,Ibrahim2021,Pogorzelski2024,Weng2026}
    &
    Integrated microwave/readout electronics and multi-site photonic sensor arrays.
    &
    Uniform sensor loading, optical coupling, and alignment across multiple channels. \\

    Microfluidic platforms
   ~\cite{sarkar2024high,10.1039/d2lc00874b,fujiwara2023diamond}
    &
    NV/ND sensing integrated with microchannels and flowing droplets.
    &
    Compact optical/MW delivery with controlled and biocompatible fluid handling. \\

    Q-BiC
   ~\cite{shanahan2025q}
    &
    Integrated microwave delivery, thermal control, and biological interfacing.
    &
    MW heating, optical access, temperature control, and biocompatibility during scaling. \\

    hBN (V$_{\mathrm{B}}^{-}$)
   ~\cite{gao2023quantum,Vlassiouk2026}
    &
    Near-surface spin defects integrated with microwave structures for liquid-phase sensing.
    &
    Reproducible, site-controlled V$_{\mathrm{B}}^{-}$ formation and biological integration. \\

    4H-SiC spin-photonic platform
   ~\cite{Lukin2020,Bao2025,Bao2026Deterministic}
    &
    Spin defects coupled to lithographically defined photonic cavities, including deterministic single-defect coupling.
    &
    Extending scalable spin--photon integration to fluidic and biological sensing. \\

    \bottomrule
    \end{tabularx}
\end{table*}

\section{Challenges and prospective solutions for optically active diamond defect sensors and fluorescent proteins}
\label{section-4}

Across single-molecule, cellular, and tissue-level NV–nanodiamond sensing studies, several intrinsic and environment-induced sensor limitations consistently emerge. At the fundamental level, sensitivity is constrained by the dipolar coupling dependence on NV–target distance, surface-induced magnetic noise, and reduced spin coherence in shallow NV centers~\cite{doi:10.1073/pnas.2509102122,https://doi.org/10.1002/adfm.202514294}. It is difficult to distinguish target-specific spin transitions from environmental magnetic noise, which restricts quantitative readout~\cite{Munuera-Javaloy2022}. Xu \textit{et al.}~\cite{doi:10.1126/sciadv.adi5300} emphasized that force sensitivity depends critically on NV depth control, calibration of spin–mechanical coupling, and suppression of background magnetic fluctuations.

When extended to cellular systems, additional constraints arise. In radical-sensing studies in granulosa cells~\cite{10.1093/hropen/hoaf007}, bronchial epithelial cells~\cite{doi:10.1021/acs.nanolett.4c02409}, and UVB-exposed keratinocytes~\cite{doi:10.1021/acssensors.4c00118}, relaxometry remains limited by nanometer-scale detection range and coherence-time reduction, while requiring careful biological controls and intracellular localization to ensure specificity. Multifunctional platforms such as the theranostic nanodiamond platform~\cite{https://doi.org/10.1002/adfm.202514294} further introduce variability due to surface functionalization effects on spin properties. In intracellular thermometry, Single-Cell Hyperthermia~\cite{https://doi.org/10.1002/adma.202517076} highlights calibration uncertainty and heating inhomogeneity as factors affecting ODMR.

At the tissue level, optical and magnetic background complexity becomes dominant.  Zhang \textit{et al}~\cite{doi:10.1073/pnas.2317921121} mentioned that optical scattering and heterogeneous background in thick liver slices can reduce contrast due to lower depth penetration and higher light attenuation, while uneven nanodiamond distribution limits proximity to dynamic biological events. de Gille \textit{et al.}~\cite{doi:10.1073/pnas.2112749118} showed that endogenous magnetic and paramagnetic sources in tissues complicate the discrimination of specific magnetic signatures in $T_1$ relaxometry. Additionally, Mzyk \textit{et al.}~\cite{https://doi.org/10.1002/adfm.202528248} reported that local chemical heterogeneity and surface interactions influence the relaxometry experiment, complicating quantitative calibration.

\begin{figure}[h]
\centering
\includegraphics[width=\linewidth]{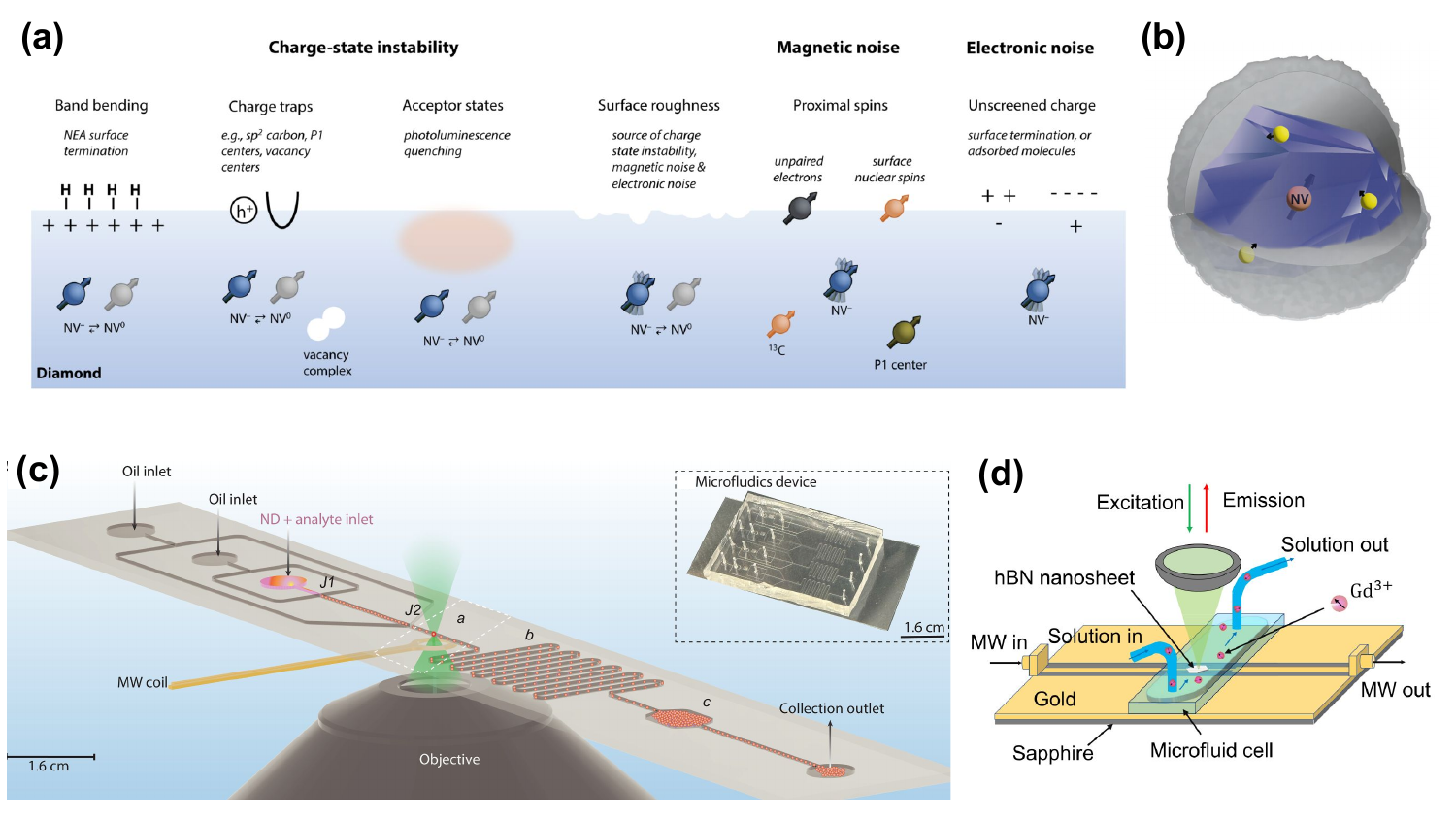}
\caption{(a) Summary of sources of charge-state instability and decoherence for near-surface NV centers. Adapted from Ref.~\cite{D2TC01258H} (b) Schematic depiction of a core-shell structured diamond nanocrystal with NV qubit (orange) in the diamond host blue. Surface spins are indicated in yellow. Adapted from Ref.~~\cite{zvi2025engineering} (c) Schematic of microfluidic chip comprising two inlets for oil and water (gray circles). $J_1$ generates ND-filled water droplets, and $J_2$ regulates droplet spacing. (region a): droplets are analyzed, (region b): additional mixing within droplets, (region c): collection chamber. (inset): photograph of the chip. Adapted from Ref.~\cite{sarkar2024high} (d) Schematic of experimental setup: an hBN nanosheet with ${V_B}^-$ spin defects is transferred onto a gold stripline microwave waveguide on a sapphire substrate. Adapted from Ref.~\cite{gao2023quantum}}
\label{S5_Fig1}
\end{figure}

NV centers in diamond have emerged as versatile quantum sensors capable of detecting several physical quantities under ambient conditions. However, the performance of shallow NV centers is strongly constrained by surface-induced instabilities (Figure~\ref{S5_Fig1}(a)). One of the main limitations is surface magnetic noise arising from fluctuating paramagnetic states, dangling bonds, residual $sp^2$ carbon, adsorbates, and surface nuclear spins, which reduce coherence and consequently limit magnetic sensitivity~\cite{D2TC01258H}.

Charge instability is another important limitation. Only the negatively charged $NV^-$ state possesses the spin-1 ground state required for quantum sensing, and its stability is strongly influenced by near-surface band bending, trap states, and surface termination. Hydrogen-terminated surfaces can induce electron depletion and destabilize $NV^-$, whereas oxygen termination generally provides improved charge stability~\cite{D2TC01258H}.

Surface functionalization therefore involves a trade-off between chemical functionality and quantum performance. In molecular sensing, the diamond surface must be modified to enable controlled attachment of target species, but additional surface layers or imperfect functionalization can introduce defect states or increase the NV--target separation, thereby weakening magnetic coupling. Janitz \textit{et al.}~\cite{D2TC01258H} therefore emphasized surface engineering as a dual requirement: the interface must support robust chemistry while preserving quantum coherence. Surface-induced decoherence in shallow NV centers has also been characterized experimentally~\cite{romach2015spectroscopy}.

Unlike solid-state hosts, proteins are structurally defined but magnetically complex, containing a dense network of nuclear spins and fluctuating local fields that inherently limit coherence times compared to solid-state defects~\cite{feder2025fluorescent,abrahams2026quantum}. Sustaining spin stability under optical excitation requires careful molecular design, as photo-induced processes and chemical heterogeneity can perturb spin populations. The maintenance of reproducible spin resonance signals across different environments is difficult due to protein folding, solvation, and variation in intracellular conditions~\cite{abrahams2026quantum}. \hl{These results highlight the need to establish response stability and reproducible calibration across different cellular environments before quantitative sensing can be extended to more complex in vivo systems.}

Yu \textit{et al.}~\cite{3j8r-2kt6} addressed surface-induced decoherence by depositing $TiO_2$ on diamond using atomic layer deposition (ALD). \hl{Following TiO$_2$ growth, the dark-spin density decreased from approximately} $2000\,\mu\mathrm{m}^{-2}$ \hl{to below the detection limit of approximately} $200\,\mu\mathrm{m}^{-2}$. \hl{For one sample after 300 ALD cycles, the Hahn-echo} $T_2$ \hl{of shallow NV centers increased by a factor of} $3.8 \pm 0.10$. The study also developed a quantitative model incorporating dark-spin density, relaxation rate, and NV-surface separation.

Biological environments introduce additional complications through charge transfer and electric-field-induced spectral shifts. \hl{Zvi \textit{et al.}}~\cite{https://doi.org/10.1002/adma.202505107} \hl{observed an average ZFS shift of $-0.45(16)$~MHz when bare nanodiamonds were immersed in phosphate-buffered saline (PBS), which they attributed to environment-induced charge redistribution and associated internal electric fields}. To stabilize the spectral response, they introduced an approximately 15-nm silica shell around the nanodiamonds (Figure~\ref{S5_Fig1}(b)), which substantially reduced ZFS drift. Surface passivation also improved spin coherence by suppressing nearby paramagnetic defects and charge traps. The $T_1$ relaxation time increased from $114\pm16~\mu$s to $379\pm33~\mu$s, while spin-echo $T_2$ increased from $1.05\pm0.41~\mu$s to $1.42\pm0.20~\mu$s. CPMG measurements further showed a 3.5-fold increase in coherence and a 3.2-fold reduction in particle-to-particle $T_2$ variation~\cite{zvi2025engineering}. Together, these studies show how surface engineering can improve both the stability and coherence of NV-based quantum sensors.

\section{Conclusion and outlook}
\label{section-5}

In conclusion, over the past decade, quantum sensing has evolved from controlled solid-state platforms to increasingly bio-integrated and hybrid systems. Advances in single-crystal diamond fabrication have enabled photonic cavities, nanomechanical resonators, and \hl{microstructured devices} that enhance \hl{spin--photon interactions} and readout efficiency~\cite{rani2020recent}. In particular, diamond micro-resonator architectures show that \hl{optical confinement and improved readout can enhance room-temperature sensitivity}~\cite{katsumi2025high}. \hl{Beyond diamond, spin defects in other materials, including optically active centers in hBN, have extended quantum sensing to alternative host systems}~\cite{roberts2025quantum,gao2023quantum}. Equally important is the shift towards biological compatibility. The \hl{Q-BiC platform shows how spin-based sensing can be integrated into a biocompatible chip for in vitro and in vivo studies}~\cite{shanahan2025q}. \hl{Microfluidic systems provide controlled biochemical environments for quantum sensing}~\cite{10.1039/d2lc00874b,fujiwara2023diamond,sarkar2024high}, while diamond nanoneedles provide a minimally invasive interface for localized biosensing~\cite{quarshie2025diamond}. \hl{Within living cells, nanodiamond quantum sensors have enabled simultaneous nanothermometry and nanorheometry}~\cite{gu2023simultaneous}, \hl{temperature measurements with a precision down to 100~mK}~\cite{sow2025millikelvin}, \hl{and quantitative viscosity mapping using NV-containing nanodiamonds}~\cite{Singam2026}. \hl{Interface chemistry remains an important factor governing coherence, stability, and sensing reliability in biological environments}~\cite{https://doi.org/10.1002/adfm.202521387}, \hl{while FND-based biosensing continues to expand toward biomedical applications}~\cite{qureshi2022recent}. \hl{Fluorescent protein spin qubits and quantum spin resonance in engineered proteins further demonstrate that optically addressable spin states can be incorporated into genetically encoded biomolecules}~\cite{abrahams2026quantum,feder2025fluorescent}. \hl{These protein-based quantum sensors complement established genetically encoded fluorescent biosensors for cellular signaling}~\cite{gest2024molecular,gonzalez2015fluorescent}. Collectively, the field now encompasses solid-state photonic devices, chip-integrated systems, intracellular nanoparticle probes, and engineered protein-based spin reporters, reflecting a shift from isolated quantum coherence \hl{demonstrations} to application-oriented sensing platforms.

Looking ahead, continued development in single-crystal diamond photonics and nanomechanics~\cite{rani2020recent,katsumi2025high} is expected \hl{to enable compact and efficient architectures. Integration with microfluidic and chip-based systems, including NV} (Figure~\ref{S5_Fig1}(c)) \hl{and hBN platforms} (Figure~\ref{S5_Fig1}(d))~\cite{sarkar2024high,gao2023quantum,shanahan2025q,10.1039/d2lc00874b,fujiwara2023diamond} \hl{can support multiplexed quantum sensing in controlled biological environments}. \hl{However, reproducible fabrication and integration of optical, microwave, electronic, and fluidic components remain important challenges for scalable implementation.} In the cellular environment, surface chemistry and functionalization strategies~\cite{https://doi.org/10.1002/adfm.202521387,zvi2025engineering,https://doi.org/10.1002/adma.202505107,qureshi2022recent} will be pivotal \hl{for probe stability, target specificity, and long-term operation}. Despite coherence times remaining limited relative to crystalline hosts, fluorescent protein-based spin systems~\cite{feder2025fluorescent,abrahams2026quantum} introduce the possibility of genetically targeted quantum sensors. \hl{The main challenge is maintaining coherence in complex biological environments.} If this challenge is successfully met, quantum optical sensing could transition from a specialized nanoscale probe \hl{toward a practical biomedical measurement technology}.


\section*{Acknowledgements} The authors are grateful to Prof. Siddharth Tallur, Prof. Abhijit Majumder, and Dr. Arpita Ghosh for their insightful thoughts.


\section*{Funding}

K.S., J.P., and A.M., acknowledge funding from DST National Quantum Mission grant number DST/FFT/NQM/QSM/2024/3.


\section*{Conflict of interest}

The authors declare no conflict of interest.




\begin{thebibliography}{10}
\newcommand{\enquote}[1]{``#1''}

\bibitem{aslam2023quantum}
N.~Aslam, H.~Zhou, E.~K. Urbach, \emph{et~al.}, \enquote{Quantum sensors for biomedical applications,} {\protect\JournalTitle{Nature Reviews Physics}} \textbf{5}, 157--169 (2023).

\bibitem{du2024single}
J.~Du, F.~Shi, X.~Kong, \emph{et~al.}, \enquote{Single-molecule scale magnetic resonance spectroscopy using quantum diamond sensors,} {\protect\JournalTitle{Reviews of Modern Physics}} \textbf{96}, 025001 (2024).

\bibitem{10.1063/5.0147469}
S.~Belser, J.~Hart, Q.~Gu, \emph{et~al.}, \enquote{Opportunities for diamond quantum metrology in biological systems,} {\protect\JournalTitle{Applied Physics Letters}} \textbf{123}, 020501 (2023).

\bibitem{degen2017quantum}
C.~L. Degen, F.~Reinhard, and P.~Cappellaro, \enquote{Quantum sensing,} {\protect\JournalTitle{Reviews of modern physics}} \textbf{89}, 035002 (2017).

\bibitem{mzyk2022relaxometry}
A.~Mzyk, A.~Sigaeva, and R.~Schirhagl, \enquote{Relaxometry with nitrogen vacancy (nv) centers in diamond,} {\protect\JournalTitle{Accounts of chemical research}} \textbf{55}, 3572--3580 (2022).

\bibitem{sengottuvel2022wide}
S.~Sengottuvel, M.~Mr{\'o}zek, M.~Sawczak, \emph{et~al.}, \enquote{Wide-field magnetometry using nitrogen-vacancy color centers with randomly oriented micro-diamonds,} {\protect\JournalTitle{Scientific reports}} \textbf{12}, 17997 (2022).

\bibitem{alam2024determining}
M.~S. Alam, F.~Gorrini, M.~Gawe{\l}czyk, \emph{et~al.}, \enquote{Determining strain components in a diamond waveguide from zero-field optically detected magnetic resonance spectra of negatively charged nitrogen-vacancy-center ensembles,} {\protect\JournalTitle{Physical Review Applied}} \textbf{22}, 024055 (2024).

\bibitem{dolde2011electric}
F.~Dolde, H.~Fedder, M.~W. Doherty, \emph{et~al.}, \enquote{Electric-field sensing using single diamond spins,} {\protect\JournalTitle{Nature Physics}} \textbf{7}, 459--463 (2011).

\bibitem{szawiola2025thermal}
G.~Szawio{\l}a, S.~Mieloch, D.~Stefa{\'n}ska, \emph{et~al.}, \enquote{Thermal effects in optically detected magnetic resonance of nitrogen vacancies in diamond: A quantum thermometer for a graduate teaching lab,} {\protect\JournalTitle{Journal of Chemical Education}} \textbf{102}, 1949--1959 (2025).

\bibitem{PhysRevB.104.224412}
K.~Zhang, S.~Ghosh, S.~Saxena, and M.~V.~G. Dutt, \enquote{Nanoscale spin detection of copper ions using double electron-electron resonance at room temperature,} {\protect\JournalTitle{Phys. Rev. B}} \textbf{104}, 224412 (2021).

\bibitem{10.1063/5.0170145}
X.~Wang, J.~Xu, S.~Ge, \emph{et~al.}, \enquote{Recent applications of nanodiamond quantum biosensors: A review,} {\protect\JournalTitle{APL Materials}} \textbf{11}, 090603 (2023).

\bibitem{schirhagl2014nitrogen}
R.~Schirhagl, K.~Chang, M.~Loretz, and C.~L. Degen, \enquote{Nitrogen-vacancy centers in diamond: nanoscale sensors for physics and biology,} {\protect\JournalTitle{Annual review of physical chemistry}} \textbf{65}, 83--105 (2014).

\bibitem{zhang2021toward}
T.~Zhang, G.~Pramanik, K.~Zhang, \emph{et~al.}, \enquote{Toward quantitative bio-sensing with nitrogen--vacancy center in diamond,} {\protect\JournalTitle{ACS sensors}} \textbf{6}, 2077--2107 (2021).

\bibitem{das2024quantum}
S.~Das, H.~Mazumdar, K.~R. Khondakar, \emph{et~al.}, \enquote{Quantum biosensors: principles and applications in medical diagnostics,} {\protect\JournalTitle{ECS Sensors Plus}} \textbf{3}, 025001 (2024).

\bibitem{chen2017diamond}
X.~Chen and W.~Zhang, \enquote{Diamond nanostructures for drug delivery, bioimaging, and biosensing,} {\protect\JournalTitle{Chemical Society Reviews}} \textbf{46}, 734--760 (2017).

\bibitem{Parhi2025.07.27.666994}
J.~Parhi, A.~Ghosh, S.~Saha, \emph{et~al.}, \enquote{Quantum sensing reveals chemotherapy induced persistent oxidative footprints in fixed brain cancer cells,} {\protect\JournalTitle{bioRxiv}}  (2026).

\bibitem{rizzato2023extending}
R.~Rizzato, M.~Schalk, S.~Mohr, \emph{et~al.}, \enquote{Extending the coherence of spin defects in hbn enables advanced qubit control and quantum sensing,} {\protect\JournalTitle{Nature communications}} \textbf{14}, 5089 (2023).

\bibitem{feder2025fluorescent}
J.~S. Feder, B.~S. Soloway, S.~Verma, \emph{et~al.}, \enquote{A fluorescent-protein spin qubit,} {\protect\JournalTitle{Nature}} \textbf{645}, 73--79 (2025).

\bibitem{abrahams2026quantum}
G.~Abrahams, A.~{\v{S}}tuhec, V.~Spreng, \emph{et~al.}, \enquote{Quantum spin resonance in engineered proteins for multimodal sensing,} {\protect\JournalTitle{Nature}} \textbf{649}, 1172--1179 (2026).

\bibitem{mzyk2021diamond}
A.~Mzyk, Y.~Ong, A.~R. Ortiz~Moreno, \emph{et~al.}, \enquote{Diamond color centers in diamonds for chemical and biochemical analysis and visualization,} {\protect\JournalTitle{Analytical chemistry}} \textbf{94}, 225--249 (2021).

\bibitem{suter2017single}
D.~Suter and F.~Jelezko, \enquote{Single-spin magnetic resonance in the nitrogen-vacancy center of diamond,} {\protect\JournalTitle{Progress in nuclear magnetic resonance spectroscopy}} \textbf{98}, 50--62 (2017).

\bibitem{levine2019principles}
E.~V. Levine, M.~J. Turner, P.~Kehayias, \emph{et~al.}, \enquote{Principles and techniques of the quantum diamond microscope,} {\protect\JournalTitle{Nanophotonics}} \textbf{8}, 1945--1973 (2019).

\bibitem{barry2020sensitivity}
J.~F. Barry, J.~M. Schloss, E.~Bauch, \emph{et~al.}, \enquote{Sensitivity optimization for nv-diamond magnetometry,} {\protect\JournalTitle{Reviews of modern physics}} \textbf{92}, 015004 (2020).

\bibitem{modak2023few}
A.~Modak, A.~Majumder, M.~Parashar, \emph{et~al.}, \enquote{Few nv center nanodiamonds enable high-speed and high-resolution sensing of paramagnetic species,} in \emph{2023 Joint Conference of the European Frequency and Time Forum and IEEE International Frequency Control Symposium (EFTF/IFCS),}  (IEEE, 2023), pp. 1--3.

\bibitem{zhu2012biocompatibilit}
Y.~Zhu, J.~Li, W.~Li, \emph{et~al.}, \enquote{The biocompatibility of nanodiamonds and their application in drug delivery systems,} {\protect\JournalTitle{Theranostics}} \textbf{2}, 302--312 (2012).

\bibitem{Müller2014}
T.~M{\"u}ller, C.~Hepp, B.~Pingault, \emph{et~al.}, \enquote{Optical signatures of silicon-vacancy spins in diamond,} {\protect\JournalTitle{Nature Communications}} \textbf{5}, 3328 (2014).

\bibitem{robertson2025photophysics}
I.~Robertson, \enquote{Photophysics and applications of spin defects in hexagonal boron nitride,} Ph.D. thesis, RMIT University (2025).

\bibitem{mendelson2022coupling}
N.~Mendelson, R.~Ritika, M.~Kianinia, \emph{et~al.}, \enquote{Coupling spin defects in a layered material to nanoscale plasmonic cavities,} {\protect\JournalTitle{Advanced Materials}} \textbf{34}, 2106046 (2022).

\bibitem{hepp2014electronic}
C.~Hepp, T.~M{\"u}ller, V.~Waselowski, \emph{et~al.}, \enquote{Electronic structure of the silicon vacancy color center in diamond,} {\protect\JournalTitle{Physical Review Letters}} \textbf{112}, 036405 (2014).

\bibitem{liu2022silicon}
W.~Liu, M.~N.~A. Alam, Y.~Liu, \emph{et~al.}, \enquote{Silicon-vacancy nanodiamonds as high performance near-infrared emitters for live-cell dual-color imaging and thermometry,} {\protect\JournalTitle{Nano letters}} \textbf{22}, 2881--2888 (2022).

\bibitem{10.1063/1.5029904}
C.~T. Nguyen, R.~E. Evans, A.~Sipahigil, \emph{et~al.}, \enquote{All-optical nanoscale thermometry with silicon-vacancy centers in diamond,} {\protect\JournalTitle{Applied Physics Letters}} \textbf{112}, 203102 (2018).

\bibitem{johnson2025nitrogen}
B.~C. Johnson, M.~O. de~Vries, A.~J. Healey, \emph{et~al.}, \enquote{The nitrogen-vacancy-nitrogen color center: A ubiquitous visible and near-infrared-ii quantum emitter in nitrogen-doped diamond,} {\protect\JournalTitle{ACS nano}} \textbf{19}, 19046--19056 (2025).

\bibitem{turcheniuk2017biomedical}
K.~Turcheniuk and V.~N. Mochalin, \enquote{Biomedical applications of nanodiamond,} {\protect\JournalTitle{Nanotechnology}} \textbf{28}, 252001 (2017).

\bibitem{10.1063/5.0202699}
Y.~A. Alzahrani and M.~Alkahtani, \enquote{Negatively charged boron-vacancy defect in hexagonal boron nitride nanoparticles,} {\protect\JournalTitle{Applied Physics Letters}} \textbf{124}, 173701 (2024).

\bibitem{Lukin2020}
D.~M. Lukin, C.~Dory, M.~A. Guidry, \emph{et~al.}, \enquote{{4H}-silicon-carbide-on-insulator for integrated quantum and nonlinear photonics,} {\protect\JournalTitle{Nature Photonics}} \textbf{14}, 330--334 (2020).

\bibitem{Bao2025}
T.~Bao, Q.~Luo, A.~Yi, \emph{et~al.}, \enquote{Tunable cavity coupling to spin defects in a {4H}-silicon-carbide-on-insulator platform,} {\protect\JournalTitle{ACS Photonics}} \textbf{12}, 2988--2996 (2025).

\bibitem{Zhou2025SiC}
Y.~Zhou, J.~Tan, H.~Hu, \emph{et~al.}, \enquote{Silicon carbide: A promising platform for scalable quantum networks,} {\protect\JournalTitle{Applied Physics Reviews}} \textbf{12}, 031301 (2025).

\bibitem{henbest2004radio}
K.~B. Henbest, P.~Kukura, C.~T. Rodgers, \emph{et~al.}, \enquote{Radio frequency magnetic field effects on a radical recombination reaction: a diagnostic test for the radical pair mechanism,} {\protect\JournalTitle{Journal of the American Chemical Society}} \textbf{126}, 8102--8103 (2004).

\bibitem{finkler2021quantum}
A.~Finkler and D.~Dasari, \enquote{Quantum sensing and control of spin-state dynamics in the radical-pair mechanism,} {\protect\JournalTitle{Physical review applied}} \textbf{15}, 034066 (2021).

\bibitem{timmel2004study}
C.~R. Timmel and K.~B. Henbest, \enquote{A study of spin chemistry in weak magnetic fields,} {\protect\JournalTitle{Philosophical Transactions of the Royal Society of London. Series A: Mathematical, Physical and Engineering Sciences}} \textbf{362}, 2573--2589 (2004).

\bibitem{yu2024noninvasive}
Z.~Yu, Y.~Xie, G.~Jin, \emph{et~al.}, \enquote{Noninvasive magnetocardiography of a living rat based on a diamond quantum sensor,} {\protect\JournalTitle{Physical Review Applied}} \textbf{21}, 064028 (2024).

\bibitem{zvi2025engineering}
U.~Zvi, D.~R. Candido, A.~M. Weiss, \emph{et~al.}, \enquote{Engineering spin coherence in core-shell diamond nanocrystals,} {\protect\JournalTitle{Proceedings of the National Academy of Sciences}} \textbf{122}, e2422542122 (2025).

\bibitem{gu2023simultaneous}
Q.~Gu, L.~Shanahan, J.~W. Hart, \emph{et~al.}, \enquote{Simultaneous nanorheometry and nanothermometry using intracellular diamond quantum sensors,} {\protect\JournalTitle{ACS nano}} \textbf{17}, 20034--20042 (2023).

\bibitem{gao2023quantum}
X.~Gao, S.~Vaidya, P.~Ju, \emph{et~al.}, \enquote{Quantum sensing of paramagnetic spins in liquids with spin qubits in hexagonal boron nitride,} {\protect\JournalTitle{ACS photonics}} \textbf{10}, 2894--2900 (2023).

\bibitem{doi:10.1021/nl502988n}
A.~O. Sushkov, N.~Chisholm, I.~Lovchinsky, \emph{et~al.}, \enquote{All-optical sensing of a single-molecule electron spin,} {\protect\JournalTitle{Nano Letters}} \textbf{14}, 6443--6448 (2014). PMID: 25333198.

\bibitem{doi:10.1021/acssensors.4c00118}
S.~Fan, L.~Lopez~Llorens, F.~P. Perona~Martinez, and R.~Schirhagl, \enquote{Quantum sensing of free radical generation in mitochondria of human keratinocytes during uvb exposure,} {\protect\JournalTitle{ACS Sensors}} \textbf{9}, 2440--2446 (2024). PMID: 38743437.

\bibitem{doi:10.1073/pnas.2112749118}
R.~W. de~Gille, J.~M. McCoey, L.~T. Hall, \emph{et~al.}, \enquote{Quantum magnetic imaging of iron organelles within the pigeon cochlea,} {\protect\JournalTitle{Proceedings of the National Academy of Sciences}} \textbf{118}, e2112749118 (2021).

\bibitem{hanlon2020diamond}
L.~Hanlon, V.~Gautam, J.~D. Wood, \emph{et~al.}, \enquote{Diamond nanopillar arrays for quantum microscopy of neuronal signals,} {\protect\JournalTitle{Neurophotonics}} \textbf{7}, 035002--035002 (2020).

\bibitem{doi:10.1126/scirobotics.aaq0076}
S.~Ghosh and A.~Ghosh, \enquote{Mobile nanotweezers for active colloidal manipulation,} {\protect\JournalTitle{Science Robotics}} \textbf{3}, eaaq0076 (2018).

\bibitem{losero2023neuronal}
E.~Losero, S.~Jagannath, M.~Pezzoli, \emph{et~al.}, \enquote{Neuronal growth on high-aspect-ratio diamond nanopillar arrays for biosensing applications,} {\protect\JournalTitle{Scientific Reports}} \textbf{13}, 5909 (2023).

\bibitem{quarshie2025diamond}
M.~Quarshie, L.~Golubewa, C.~Giraulo, \emph{et~al.}, \enquote{Diamond nanoneedles for biosensing,} {\protect\JournalTitle{Nanotechnology}} \textbf{36}, 165501 (2025).

\bibitem{doi:10.1073/pnas.2509102122}
M.~Li, Q.~Zhang, X.~Kong, \emph{et~al.}, \enquote{Quantum relaxometry for detecting biomolecular interactions with single nv centers,} {\protect\JournalTitle{Proceedings of the National Academy of Sciences}} \textbf{122}, e2509102122 (2025).

\bibitem{Munuera-Javaloy2022}
C.~Munuera-Javaloy, R.~Puebla, B.~D'Anjou, \emph{et~al.}, \enquote{Detection of molecular transitions with nitrogen-vacancy centers and electron-spin labels,} {\protect\JournalTitle{npj Quantum Information}} \textbf{8}, 140 (2022).

\bibitem{doi:10.1126/sciadv.adi5300}
F.~Xu, S.~Zhang, L.~Ma, \emph{et~al.}, \enquote{Quantum-enhanced diamond molecular tension microscopy for quantifying cellular forces,} {\protect\JournalTitle{Science Advances}} \textbf{10}, eadi5300 (2024).

\bibitem{Bray2024}
F.~Bray, M.~Laversanne, H.~Sung, \emph{et~al.}, \enquote{Global cancer statistics 2022: Globocan estimates of incidence and mortality worldwide for 36 cancers in 185 countries,} {\protect\JournalTitle{CA: A Cancer Journal for Clinicians}} \textbf{74}, 229--263 (2024).

\bibitem{zhang2025global}
Y.~Zhang, Y.~Ji, S.~Liu, \emph{et~al.}, \enquote{Global burden of female breast cancer: new estimates in 2022, temporal trend and future projections up to 2050 based on the latest release from globocan,} {\protect\JournalTitle{Journal of the National Cancer Center}} \textbf{5}, 287--296 (2025).

\bibitem{https://doi.org/10.1002/adfm.202514294}
K.~Wu, S.~Fan, Y.~Zhang, \emph{et~al.}, \enquote{Fluorescent nanodiamonds based theranostic platform for ph-sensitive drug delivery and quantum sensing,} {\protect\JournalTitle{Advanced Functional Materials}} \textbf{36}, e14294 (2026).

\bibitem{10.1093/hropen/hoaf007}
N.~Lin, K.~C. van Zomeren, T.~Plosch, \emph{et~al.}, \enquote{Follicle-stimulating hormone stimulates free radical generation without inducing substantial oxidative stress in human granulosa cells,} {\protect\JournalTitle{Human Reproduction Open}} \textbf{2025}, hoaf007 (2025).

\bibitem{doi:10.1021/acs.nanolett.4c02409}
Y.~Zhang, A.~Sigaeva, S.~Fan, \emph{et~al.}, \enquote{Dynamics for high-sensitivity detection of free radicals in primary bronchial epithelial cells upon stimulation with cigarette smoke extract,} {\protect\JournalTitle{Nano Letters}} \textbf{24}, 9650--9657 (2024). PMID: 39012318.

\bibitem{https://doi.org/10.1002/adma.202517076}
K.~Wu, Q.~Lu, Y.~Ren, \emph{et~al.}, \enquote{Single-cell hyperthermia: Diamond quantum thermometry reveals thermal control of macrophage polarization,} {\protect\JournalTitle{Advanced Materials}} \textbf{38}, e17076 (2026).

\bibitem{10.1021/acsanm.5c05655}
J.~C. Consiglio, R.~Boltyanskiy, Y.~L. Mindarava, \emph{et~al.}, \enquote{Optically detected magnetic resonance based intracellular thermometry using nanodiamonds implanted in adherent cancer cells,} {\protect\JournalTitle{ACS Applied Nano Materials}} \textbf{9}, 3590--3596 (2026).

\bibitem{sow2025millikelvin}
M.~Sow, J.~Mohnani, G.~Genov, \emph{et~al.}, \enquote{Millikelvin intracellular nanothermometry with nanodiamonds,} {\protect\JournalTitle{Advanced Science}} \textbf{12}, e11670 (2025).

\bibitem{Singam2026}
A.~Singam, C.~Lee, D.~Kim, \emph{et~al.}, \enquote{Holotomography-based quantification of intracellular viscosity using nanodiamonds containing nitrogen-vacancy (nv) centers,} {\protect\JournalTitle{ACS Applied Nano Materials}} \textbf{9}, 16205--16213 (2026).

\bibitem{doi:10.1073/pnas.2317921121}
Y.~Zhang, A.~Sigaeva, A.~Elías-Llumbet, \emph{et~al.}, \enquote{Free radical detection in precision-cut mouse liver slices with diamond-based quantum sensing,} {\protect\JournalTitle{Proceedings of the National Academy of Sciences}} \textbf{121}, e2317921121 (2024).

\bibitem{https://doi.org/10.1002/adfm.202528248}
A.~Mzyk, E.~Yilmaz, G.~D. Angelo, and K.~Berg-Sørensen, \enquote{Nanodiamond-based quantum sensing reveals changes in endo-lysosomal activity of stimulated cardiac fibroblasts in a cellular fibrosis model,} {\protect\JournalTitle{Advanced Functional Materials}} \textbf{36}, e28248 (2026).

\bibitem{doi:10.1021/acs.nanolett.1c03021}
L.~Nie, A.~C. Nusantara, V.~G. Damle, \emph{et~al.}, \enquote{Quantum sensing of free radicals in primary human dendritic cells,} {\protect\JournalTitle{Nano Letters}} \textbf{22}, 1818--1825 (2022). PMID: 34929080.

\bibitem{doi:10.1073/pnas.1601513113}
J.~F. Barry, M.~J. Turner, J.~M. Schloss, \emph{et~al.}, \enquote{Optical magnetic detection of single-neuron action potentials using quantum defects in diamond,} {\protect\JournalTitle{Proceedings of the National Academy of Sciences}} \textbf{113}, 14133--14138 (2016).

\bibitem{Bathla:26}
A.~Bathla, M.~Parashar, M.~Markham, \emph{et~al.}, \enquote{Dictionary-based reconstruction of spatio-temporal 3d magnetic field images from a quantum diamond microscope,} {\protect\JournalTitle{Optica}} \textbf{13}, 686--697 (2026).

\bibitem{Hamoya2024}
T.~Hamoya, K.~Kaminaga, R.~Igarashi, \emph{et~al.}, \enquote{Intravital microscopic thermometry of rat mammary epithelium by fluorescent nanodiamond,} {\protect\JournalTitle{Nanoscale Horizons}} \textbf{9}, 1938--1947 (2024).

\bibitem{shanahan2025q}
L.~Shanahan, S.~Belser, J.~W. Hart, \emph{et~al.}, \enquote{Q-bic: a biocompatible integrated chip for in vitro and in vivo spin-based quantum sensing,} {\protect\JournalTitle{PRX Life}} \textbf{3}, 013016 (2025).

\bibitem{Shimomura2026}
S.~Shimomura, Y.~Nakane, H.~Abe, \emph{et~al.}, \enquote{Defect-activated and surface-modified hexagonal boron nitride nanoparticles toward intracellular quantum sensing,} {\protect\JournalTitle{Nano Letters}} \textbf{26}, 3674--3680 (2026).

\bibitem{Vlassiouk2026}
I.~V. Vlassiouk, Y.-C. Wu, A.~Puretzky, \emph{et~al.}, \enquote{Defect engineering in large-scale {CVD}-grown hexagonal boron nitride: Formation, spectroscopy, and spin relaxation dynamics,} {\protect\JournalTitle{Small}} \textbf{22}, e06874 (2026).

\bibitem{Sarkar2026Dual}
S.~Sarkar, N.~Agrawal, S.~Dasika, and K.~Saha, \enquote{Real-time simultaneous dual sensing of temperature and magnetic field using nitrogen-vacancy based nanodiamonds,} {\protect\JournalTitle{Measurement}} \textbf{257}, 118462 (2026).

\bibitem{Santagati2019}
R.~Santagati, A.~A. Gentile, S.~Knauer, \emph{et~al.}, \enquote{Magnetic-field learning using a single electronic spin in diamond with one-photon readout at room temperature,} {\protect\JournalTitle{Physical Review X}} \textbf{9}, 021019 (2019).

\bibitem{zhu2023multicone}
T.~Zhu, J.~Rhensius, K.~Herb, \emph{et~al.}, \enquote{Multicone diamond waveguides for nanoscale quantum sensing,} {\protect\JournalTitle{Nano Letters}} \textbf{23}, 10110--10117 (2023).

\bibitem{rani2020recent}
D.~Rani, O.~R. Opaluch, and E.~Neu, \enquote{Recent advances in single crystal diamond device fabrication for photonics, sensing and nanomechanics,} {\protect\JournalTitle{Micromachines}} \textbf{12}, 36 (2020).

\bibitem{shukla2023opto}
A.~Shukla, S.~Tiwari, A.~Majumder, \emph{et~al.}, \enquote{Opto-thermoelectric trapping of fluorescent nanodiamonds on plasmonic nanostructures,} {\protect\JournalTitle{Optics Letters}} \textbf{48}, 2937--2940 (2023).

\bibitem{Kim2019}
D.~Kim, M.~I. Ibrahim, C.~Foy, \emph{et~al.}, \enquote{A {CMOS}-integrated quantum sensor based on nitrogen--vacancy centres,} {\protect\JournalTitle{Nature Electronics}} \textbf{2}, 284--289 (2019).

\bibitem{Ibrahim2021}
M.~I. Ibrahim, C.~Foy, D.~R. Englund, and R.~Han, \enquote{High-scalability {CMOS} quantum magnetometer with spin-state excitation and detection of diamond color centers,} {\protect\JournalTitle{IEEE Journal of Solid-State Circuits}} \textbf{56}, 1001--1014 (2021).

\bibitem{Pogorzelski2024}
J.~Pogorzelski, L.~Horsthemke, J.~Homrighausen, \emph{et~al.}, \enquote{Compact and fully integrated {LED} quantum sensor based on {NV} centers in diamond,} {\protect\JournalTitle{Sensors}} \textbf{24}, 743 (2024).

\bibitem{Weng2026}
H.-C. Weng, J.~G. Rarity, K.~C. Balram, and J.~A. Smith, \enquote{Photonic-integrated quantum sensor array for microscale magnetic localisation,} {\protect\JournalTitle{Nature Communications}} \textbf{17}, 6185 (2026).

\bibitem{sarkar2024high}
A.~Sarkar, Z.~R. Jones, M.~Parashar, \emph{et~al.}, \enquote{High-precision chemical quantum sensing in flowing monodisperse microdroplets,} {\protect\JournalTitle{Science Advances}} \textbf{10}, eadp4033 (2024).

\bibitem{10.1039/d2lc00874b}
R.~D. Allert, F.~Bruckmaier, N.~R. Neuling, \emph{et~al.}, \enquote{Microfluidic quantum sensing platform for lab-on-a-chip applications,} {\protect\JournalTitle{Lab on a Chip}} \textbf{22}, 4831--4840 (2022).

\bibitem{fujiwara2023diamond}
M.~Fujiwara, \enquote{Diamond quantum sensors in microfluidics technology,} {\protect\JournalTitle{Biomicrofluidics}} \textbf{17} (2023).

\bibitem{Bao2026Deterministic}
T.~Bao, Q.~Luo, A.~Yi \emph{et~al.}, \enquote{Deterministic and scalable coupling of single {4H-SiC} spin defects into bullseye cavities,} {\protect\JournalTitle{Laser \& Photonics Reviews}} \textbf{20}, e02377 (2026).

\bibitem{D2TC01258H}
E.~Janitz, K.~Herb, L.~A. Völker, \emph{et~al.}, \enquote{Diamond surface engineering for molecular sensing with nitrogen—vacancy centers,} {\protect\JournalTitle{J. Mater. Chem. C}} \textbf{10}, 13533--13569 (2022).

\bibitem{romach2015spectroscopy}
Y.~Romach, C.~M{\"u}ller, T.~Unden, \emph{et~al.}, \enquote{Spectroscopy of surface-induced noise using shallow spins in diamond,} {\protect\JournalTitle{Physical review letters}} \textbf{114}, 017601 (2015).

\bibitem{3j8r-2kt6}
X.~Yu, E.~J. Villafranca, S.~Wang, \emph{et~al.}, \enquote{Engineering diamond interfaces free of dark spins,} {\protect\JournalTitle{Phys. Rev. Appl.}} \textbf{25}, 034006 (2026).

\bibitem{https://doi.org/10.1002/adma.202505107}
U.~Zvi, S.~Mundhra, D.~Ovetsky, \emph{et~al.}, \enquote{Probing cellular activity via charge-sensitive quantum nanoprobes,} {\protect\JournalTitle{Advanced Materials}} \textbf{38}, e05107 (2026).

\bibitem{katsumi2025high}
R.~Katsumi, K.~Takada, K.~Kawai, \emph{et~al.}, \enquote{High-sensitivity nanoscale quantum sensors based on a diamond micro-resonator,} {\protect\JournalTitle{Communications Materials}} \textbf{6}, 49 (2025).

\bibitem{roberts2025quantum}
H.~Roberts, H.~Abudayyeh, X.~Li, and X.~Li, \enquote{Quantum sensing with spin defects beyond diamond,} {\protect\JournalTitle{ACS nano}} \textbf{19}, 22528--22575 (2025).

\bibitem{https://doi.org/10.1002/adfm.202521387}
A.~Sadžak, I.~Pérez, M.~Guimarães, \emph{et~al.}, \enquote{Impact of surface functionalization on nv quantum properties: Implications for biosensing with fluorescent nanodiamonds,} {\protect\JournalTitle{Advanced Functional Materials}} \textbf{36}, e21387 (2026).

\bibitem{qureshi2022recent}
S.~A. Qureshi, W.~W.-W. Hsiao, L.~Hussain, \emph{et~al.}, \enquote{Recent development of fluorescent nanodiamonds for optical biosensing and disease diagnosis,} {\protect\JournalTitle{Biosensors}} \textbf{12}, 1181 (2022).

\bibitem{gest2024molecular}
A.~M. Gest, A.~Z. Sahan, Y.~Zhong, \emph{et~al.}, \enquote{Molecular spies in action: Genetically encoded fluorescent biosensors light up cellular signals,} {\protect\JournalTitle{Chemical Reviews}} \textbf{124}, 12573--12660 (2024).

\bibitem{gonzalez2015fluorescent}
J.~A. Gonz{\'a}lez-Vera and M.~C. Morris, \enquote{Fluorescent reporters and biosensors for probing the dynamic behavior of protein kinases,} {\protect\JournalTitle{Proteomes}} \textbf{3}, 369--410 (2015).

\end{thebibliography}



\end{document}